# The Influence of Central Body Tides on Catastrophic Disruptions of Close-in Planetary Satellites

Harrison Agrusa[1] and Patrick Michel[1,2]  
[1] Université Côte d'Azur, Observatoire de la Côte d'Azur, CNRS, Laboratoire Lagrange, Nice, France; hagrusa@oca.eu  
[2] The University of Tokyo, Department of Systems Innovation, School of Engineering, Tokyo, Japan

## Abstract

We model the outcomes of catastrophic disruptions on small, gravity-dominated natural satellites, accounting for the tidal potential of the central body, which is neglected in classical disruption scaling laws. We introduce the concept of $Q_{\rm TD}^\star$, the specific energy required to disperse half of the total mass involved in a collision, accounting for the tidal potential of a central body. We derive a simple scaling relation for $Q_{\rm TD}^\star$ and demonstrate that for close-in planetary or asteroidal satellites, the tides from the central body can significantly reduce their catastrophic disruption threshold. We show that many satellites in the solar system are in such a regime, where their disruption threshold should be much lower than that predicted by classical scaling laws that neglect tidal effects. Some notable examples include Mars's Phobos, Jupiter's Metis and Adrastea, Saturn's ring moons, Uranus's Ophelia, and Neptune's Naiad and Thalassa, among others. We argue that traditional impact scaling laws should be modified to account for tides when modeling the formation and evolution of these close-in satellites. Our derivation for $Q_{\rm TD}^\star$ can easily be used in existing N-body and collisional evolution codes.

*Unified Astronomy Thesaurus concepts:* Collisional processes (2286); Natural satellite evolution (2297); Tides (1702); Small Solar System bodies (1469)

## 1. Introduction

Impact processes play a dominant role in solar system evolution at nearly all size scales, including the growth of planetesimals (A. Johansen et al. 2014), terrestrial planet formation (C. B. Agnor et al. 1999; J. E. Chambers 2013), and the collisional evolution of small-body populations (W. F. Bottke et al. 2005, 2015). The outcomes of collisions are often parameterized by the specific impact energy, or $Q$. In particular, the catastrophic disruption threshold, $Q_{\rm D}^\star$, is defined as the specific energy required to disperse half of the total mass involved in a collision. Numerically, it has been shown that $Q_{\rm D}^\star$ is a complicated function of the target's physical and material properties (size, density, spin, material properties, etc.) and the impact conditions (impactor size, velocity, etc.; e.g., W. Benz & E. Asphaug 1999; M. Jutzi et al. 2010; Z. M. Leinhardt & S. T. Stewart 2012; R.-L. Ballouz et al. 2014; M. Jutzi 2015; S. D. Raducan et al. 2024). Scaling laws for $Q_{\rm D}^\star$ are often used each time a diagnostic needs to be made on a collision outcome in various contexts without needing to directly simulate the collision itself.

To apply this concept of $Q_{\rm D}^\star$ to natural satellites, we consider the role of an external tidal potential on the outcomes of large collisions on gravity-dominated bodies. R. Hyodo & K. Ohtsuki (2014) demonstrated that Saturn's tides can affect the disruption threshold for close-in moonlets. We seek to quantify this effect further and develop a scaling law that can be applied to any gravity-dominated body feeling a strong tidal potential. We introduce the concept of $Q_{\rm TD}^\star$, which accounts for the influence of tides on the catastrophic disruption threshold. For close-in planetary satellites, we show that tides substantially reduce the catastrophic disruption threshold and may play a significant role in their collisional evolution. We provide a simple derivation for $Q_{\rm TD}^\star$, which can easily be included in N-body codes with prescription-based treatments of catastrophic disruptions. Understanding the influence of tides on collision outcomes may have strong implications for the origins, histories, and lifetimes of some of the solar system's numerous close-in natural satellites.

In Section 2, we briefly introduce some background material based on existing literature. Section 3 introduces the numerical methods and simulation setup. The results are presented in Section 4, along with some discussion in Section 5. Finally, some conclusions and perspectives are in Section 6.

## 2. Background

We follow a similar convention to that of S. T. Stewart & Z. M. Leinhardt (2009) and Z. M. Leinhardt & S. T. Stewart (2012) by writing the specific impact energy in terms of the reduced mass, $\mu$, which we denote using the subscript "R." The specific impact energy of a collision between a projectile with mass $M_{\rm proj}$ and a target with mass $M_{\rm targ}$ is then defined as

$$Q_{\rm R} = \frac{1}{2}\frac{\mu}{M_{\rm tot}}v_{\rm imp}^2, \qquad (1)$$

where $v_{\rm imp}$ is the impact speed, $M_{\rm tot} = M_{\rm targ} + M_{\rm proj}$, and $\mu = M_{\rm proj}M_{\rm targ}/(M_{\rm targ} + M_{\rm proj})$. Previous numerical studies have demonstrated that the outcomes of gravity-dominated collisions tend to follow what they call the "universal law" for the mass of the largest remnant (Z. M. Leinhardt & S. T. Stewart 2009; S. T. Stewart & Z. M. Leinhardt 2009). That is, the mass of the largest remnant following a catastrophic collision tends to decrease linearly with increasing specific impact energy. For head-on impacts, this can be

written as

$$\frac{M_{\mathrm{lr}}}{M_{\mathrm{tot}}} = -\frac{1}{2}\left(\frac{Q_R}{Q_{\mathrm{RD}}^\star} - 1\right) + \frac{1}{2}, \quad (2)$$

where $M_{\mathrm{lr}}$ is the mass of the largest remnant and $Q_{\mathrm{RD}}^\star$ is the catastrophic disruption threshold, defined to be the specific impact energy such that $M_{\mathrm{lr}} = 0.5 M_{\mathrm{tot}}$ (Z. M. Leinhardt & S. T. Stewart 2009). $Q_{\mathrm{RD}}^\star$ can either be computed using scaling laws (e.g., K. R. Housen & K. A. Holsapple 1990) or computed by fitting the results of numerical simulations to Equation (2) (e.g., Z. M. Leinhardt & S. T. Stewart 2012). Despite its nickname, the so-called "universal law" is not always universal, as it tends to break down at impact energies significantly below or above $Q_D^\star$ and in other special cases (e.g., R.-L. Ballouz et al. 2015; C. Reinhardt et al. 2022; S. Crespi et al. 2024).

R. Hyodo & K. Ohtsuki (2014) considered the outcomes of catastrophic collisions for gravitational aggregates around Saturn and found that the so-called "universal law" breaks down for orbital distances less than ∼200,000 km (∼3.4 Saturn radii). However, this study only considered equal-mass collisions around a single planet. Inspired by this study and the potential implications for the collisional lifetime of planetary satellites, we seek to develop a more general model that can be applied to any planetary or asteroid system with close-in satellites. In order to generalize our results, we adopt normalized units for the target's orbital distance and spin rate, which allow the results to easily be applied to any system (K. A. Holsapple & P. Michel 2006). We define the normalized orbital distance, $\delta$, and normalized spin rate $\Omega$, as

$$\delta = \left(\frac{\rho}{\rho_P}\right)^{1/3} \frac{d}{R_P} \quad (3)$$

$$\Omega = \frac{\omega}{\sqrt{\pi \rho G}}, \quad (4)$$

where $\rho$ is the target (satellite) bulk density, $\rho_P$ and $R_P$ are the primary's (planet) respective bulk density and radius, $d$ is the orbital distance, $\omega$ is the spin rate, and $G$ is the standard gravitational constant. Recall that for a uniform-density sphere, the spin limit, where the centrifugal acceleration is balanced by self-gravity, is $\omega_{\mathrm{crit}} = \sqrt{4\pi\rho G/3}$, corresponding to $\Omega_{\mathrm{crit}} = \sqrt{4/3}$. In addition, for a tidally locked satellite, these two normalized parameters can be simply related through Kepler's third law: $\Omega^2 = 4\delta^{-3}/3$. We also note that the tidal disruption distance is $\delta_{\mathrm{Roche}} \sim 1.5$ based on the Drucker–Prager failure criterion for a spherical, uniform-density, cohesionless rubble pile with an angle of internal friction of ∼35°, which is a typical angle for ordinary granular material. This distance varies as a function of the satellite's shape and material properties, but $\delta_{\mathrm{Roche}} \sim 1.5$ is a useful number to keep in mind.

## 3. Methods

We use PKDGRAV, a gravitational $N$-body and granular dynamics code (D. C. Richardson et al. 2000; S. R. Schwartz et al. 2014; Y. Zhang et al. 2017) to model the outcomes of low-speed collisions between gravity-dominated bodies. The addition of tides in collisional disruptions removes the spherical symmetry of the problem and significantly increases the number of free parameters. In order to keep the problem computationally tractable, we consider targets with a single set of physical and material properties and only test two unique impact geometries.

In all simulations, the target consists of $10^4$ randomly arranged particles following a differential power-law size-frequency distribution with a slope of −3, truncated between radii of 250 and 750 m. Particles are cohesionless, and the friction parameters are tuned such that the body has a friction angle of ∼35° (Y. Zhang et al. 2022). A friction angle of 35° was chosen because it is a typical angle of repose for granular materials (J. K. Mitchell et al. 1972; C. A. Bareither et al. 2008), and it is similar to the friction angle estimates for several recently visited asteroids (A. Fujiwara et al. 2006; O. Barnouin et al. 2019, 2024; S. Watanabe et al. 2019; C. Q. Robin et al. 2024) The particle density is set to 2 g cm$^{-3}$, which corresponds to a target bulk density of $\rho_{\mathrm{bulk}} \sim 1.35$ g cm$^{-3}$ and a bulk radius of $R \sim 10$ km. The target escape speed is $v_{\mathrm{esc}} \sim 9$ m s$^{-1}$.

The target is placed on a circular orbit around a central massive particle that provides the tidal potential. The target is settled into an equilibrium before the impactor is introduced in the simulation and collides with the target. For simplicity, the impactor has the same physical and material properties as the target. To reduce the number of free parameters, it is assumed that the target is tidally locked to the primary, meaning that the target's spin rate is not held constant and is instead a function of its orbital distance, while the impactor has no rotation. In addition, all impacts are head-on, although we test two impact geometries: one where the impactor is coming from the direction of the primary (i.e., the negative radial direction denoted -R) and the other where it is coming from the direction of the target's orbital velocity (i.e., the positive tangential direction denoted +T). These two geometries are shown in Figure 1. Our simulations vary the orbital distance at which the impact occurs as well as the impactor's size and velocity, while keeping all other impactor and target parameters fixed. Due to the asymmetry introduced by the tidal potential, the impact direction can play a crucial role, despite the impact angle being fixed (R. Hyodo & K. Ohtsuki 2014). The orbital distance is varied between $\delta = 1.7$ and $\delta = 10$. Inside of $\delta = 1.7$, the target oftentimes tidally disrupts prior to the impact, while outside $\delta = 10$, the strength is negligibly small.

All simulations are run for $3 \times 10^7$ time steps, which correspond to ∼42 hr. This simulation time was determined by eye, where most simulations have reached a terminal end state in terms of the mass of the largest remnant. For the closest ($\delta = 1.7$) and most distant impacts ($\delta = 10$), this corresponds to ∼6.6 and ∼0.5 orbit periods, respectively. A total of 540 simulations were performed with varying impactor masses, speeds, and orbital distances, which are tabulated in Appendix A.

## 4. Results

### 4.1. Equal-mass Collisions

First, we show some outcomes of equal-mass collisions to demonstrate some basic findings. In Figure 2, we show time-series plots for equal-mass collisions for various impact speeds

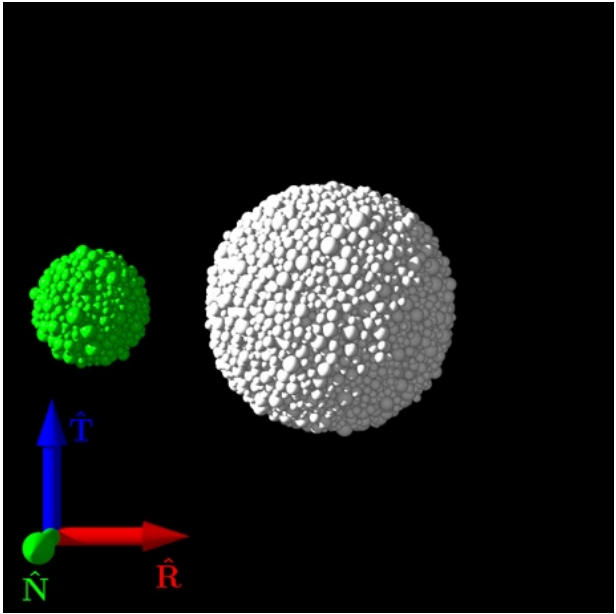
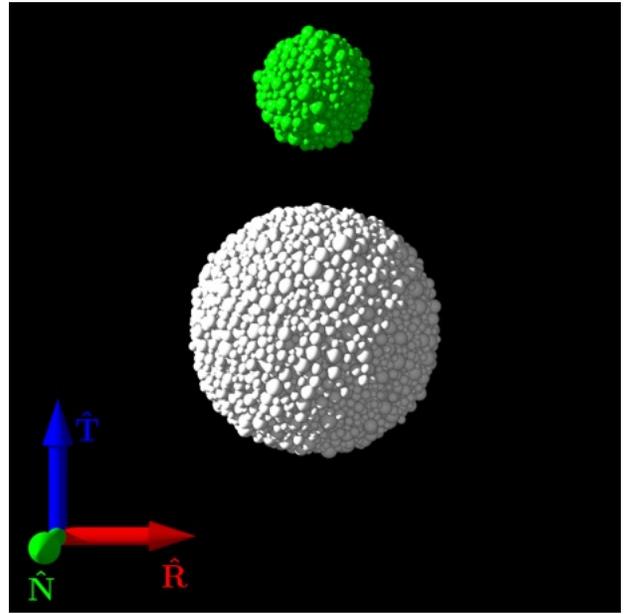

(a) An impact coming from the -$\hat{R}$ direction.

(b) An impact coming from the +$\hat{T}$ direction.

**Figure 1.** The two impact geometries considered in this study, shown in a radial–tangential–normal (RTN) coordinate frame. The $\hat{R}$ direction points directly away from the central body, $\hat{T}$ points along the target's orbital velocity vector, and $\hat{N}$ completes the right-handed triad, pointing along the orbit normal. For a "-R" collision, the impactor comes from the $-\hat{R}$ direction with a relative velocity in the $+\hat{R}$ direction. Similarly, "+T" collisions have the impactor coming from the $+\hat{T}$ direction with its relative velocity in the $-\hat{T}$ direction.

for each orbital distance. At the low end, when $v_{\text{imp}}/v_{\text{esc}} = 1$, the collisions are extremely gentle, resulting in either a perfect merger for distant collisions or hit-and-runs for close-in collisions. The hit-and-runs are the result of the collision forming a temporary contact binary, which is then ripped apart as a result of the strong tides, which is shown in Figure 3. We note that this could be an important constraint for forming contact binaries like Selam, which is the satellite of the Dinkinesh binary system discovered during the Dinkinesh flyby by the NASA Lucy mission (H. F. Levison et al. 2024). Because the two lobes of Selam are large relative to Dinkinesh, their orbital speeds are comparable to their surface escape speeds, meaning a collision between the two lobes of Selam will occur around $v_{\text{imp}}/v_{\text{esc}} \sim 1$. Here, we are only testing head-on impacts and a single collision geometry, so a more detailed study would be needed to constrain the origin of Selam, although these simulations illustrate that outcomes will be sensitive to the orbital distance of a potential Selam-forming merger.

When the impact speed is increased slightly to $v_{\text{imp}} = 2v_{\text{esc}}$, Figure 2 shows that the outcomes become extremely bimodal, resulting in either a perfect merger or a complete disruption of the target. Sometimes, the disruptions do not occur until up to $\sim$10 hr postmerger, as a result of the collision triggering a tidal disruption. Some simulation snapshots are shown in Figure 4 for a fixed collision speed of $v_{\text{imp}}/v_{\text{esc}} = 2$ at three orbital distances corresponding to 2.0, 2.2, and 10.0 in normalized units of $\delta$. Taking a close look at the case with $\delta = 2$, we see that after $\sim$4 hr, the collision seems to have resulted in a perfect merger. However, several hours later, the combined target and projectile undergo a complete tidal disruption. If the distance is increased slightly to $\delta = 2.2$, the first several panels of Figure 4 look nearly identical. This time, however, the target avoids a tidal disruption and successfully merges. We see a similar outcome when the distance is increased substantially to $\delta = 10$, although the resulting shape is uniquely different, which we attribute to the weaker tidal potential allowing the body to maintain a more irregular, ravioli-like shape. At $v_{\text{imp}} = 4v_{\text{esc}}$, we see similar behavior and bimodal outcomes. Finally, at $v_{\text{imp}} = 6v_{\text{esc}}$, almost all collisions are catastrophic, with the close-in collisions being supercatastrophic as tides prevent them from easily reaccreting any material.

### 4.2. Low-mass Ratio Collisions

The lowest-mass ratio collisions considered here are $M_{\text{proj}}/M_{\text{targ}} = 0.01$, as going any lower requires supersonic impact speeds to achieve disruptions, which is a regime where a hydrocode would be needed to model the relevant physics, including shock waves or vaporization (M. Jutzi et al. 2015). In Figure 5, we show snapshots from three simulations with $M_{\text{proj}}/M_{\text{targ}} = 0.01$, an impact speed of $v_{\text{imp}} = 30v_{\text{esc}}$, and the radial (-R) impact geometry at three unique normalized orbit distances: $\delta = 1.8, 2.4,$ and $10$. Despite having the same specific impact energies, these three collisions result in very different outcomes; when $\delta = 1.8$, the collision immediately leads to a tidal disruption, effectively making the collision supercatastrophic. When the distance is increased slightly to $\delta = 2.4$, the target is able to reaccrete some material, and at the 8 hr mark, the largest remnant is about 80% of the total mass. However, this collision significantly disturbs the shape, rotation state, and orbit of the target, which ultimately leads to a series of small-mass stripping events starting 16 hr postcollision, resulting in an additional mass loss of $\sim$5%. Finally, when $\delta = 10$, the tidal potential and target's synchronous rotation play only a very small role in the collision outcome, and we end with $M_{\text{lr}}/M_{\text{tot}} \sim 0.8$.

In Figure 6, we show several time-series plots of $M_{\text{lr}}/M_{\text{tot}}$ for various impact speeds and orbital distances when $M_{\rm proj}/M_{\rm targ} = 0.01$. These plots highlight the bimodal nature of collision outcomes when tides are strong: collisions typically result in either a very large or very small largest remnant. Even for relatively small impactors, impact speeds as low as $v_{\rm imp}/v_{\rm esc} = 2$ can lead to complete tidal disruptions when the target is already close to the Roche limit. They also demonstrate the time-dependent nature of the outcome, as many cases result in a complete tidal disruption of the target many hours after the collision occurs. In fact, in some cases, the target begins tidally disrupting toward the end of the simulation and would likely be completely destroyed if the simulations were run longer. In these cases, the disruption occurs several orbit periods after the impact. The root cause is the impact reshaping the target and placing it on an eccentric orbit with an excited rotation state, eventually leading to a tidal disruption. If the simulations were run longer, it is likely that several other cases would have resulted in complete disruptions,
meaning that the resulting estimates of $Q_D^\star$ could arguably be considered an underestimate.

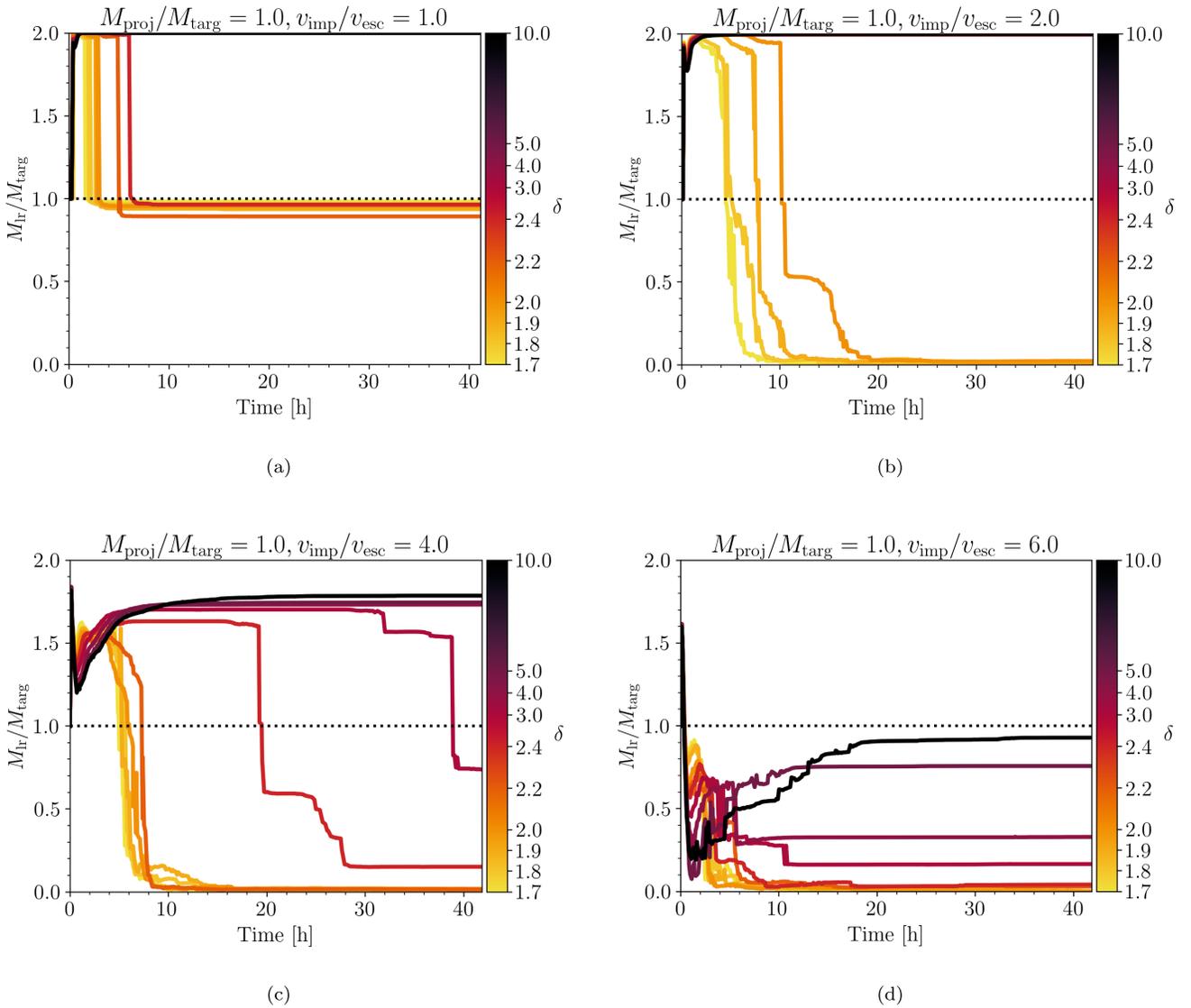

**Figure 2.** Time-series plots for equal-mass impacts for fixed impact speeds and varying distances with the -R impact geometry. Each plot shows the ratio of the largest remnant to the target mass as a function of time (in hours) for a specific impact speed, indicated on the top of each subfigure. The color scale indicates the distance to the central body in normalized units. Plotting $M_{\rm lr}$ normalized to the target mass, rather than the total mass, makes it easier to distinguish the threshold between an erosive and accretionary collision, which is indicated by the dotted black line.

### 4.3. Fitting Impact Outcomes to the "Universal Law"

In Figure 7, we show the results of fitting all simulations with the -R impact geometry and a $M_{\rm proj}/M_{\rm targ}$ of 0.1 to the "universal law" (Equation (2)). Here, for each unique orbital distance ($\delta$), we perform a nonlinear least-squares fit to determine the value of $Q_{\rm RD}^\star$ that best places the resulting $M_{\rm lr}/M_{\rm tot}$ along the black line representing the universal law. The only parameter that varies in each fit is the impact speed (i.e., $Q_R$). For $\delta \gtrsim 3$, we see that the universal law does a reasonably good job of predicting the mass of the largest remnant, as we see a nice linear trend between $M_{\rm lr}$ and $Q_R$. Below this distance, however, it seems that the universal law breaks down. In only the lowest-energy impacts do we see that $M_{\rm lr}/M_{\rm tot}$ remains close to unity. Then, as $Q_R$ is increased slightly, we see an abrupt drop in the resulting $M_{\rm lr}$. This is consistent with the findings of R. Hyodo & K. Ohtsuki (2014), who found that the universal law begins to break down inside of ~3.4 Saturn radii for head-on, equal-mass collisions.

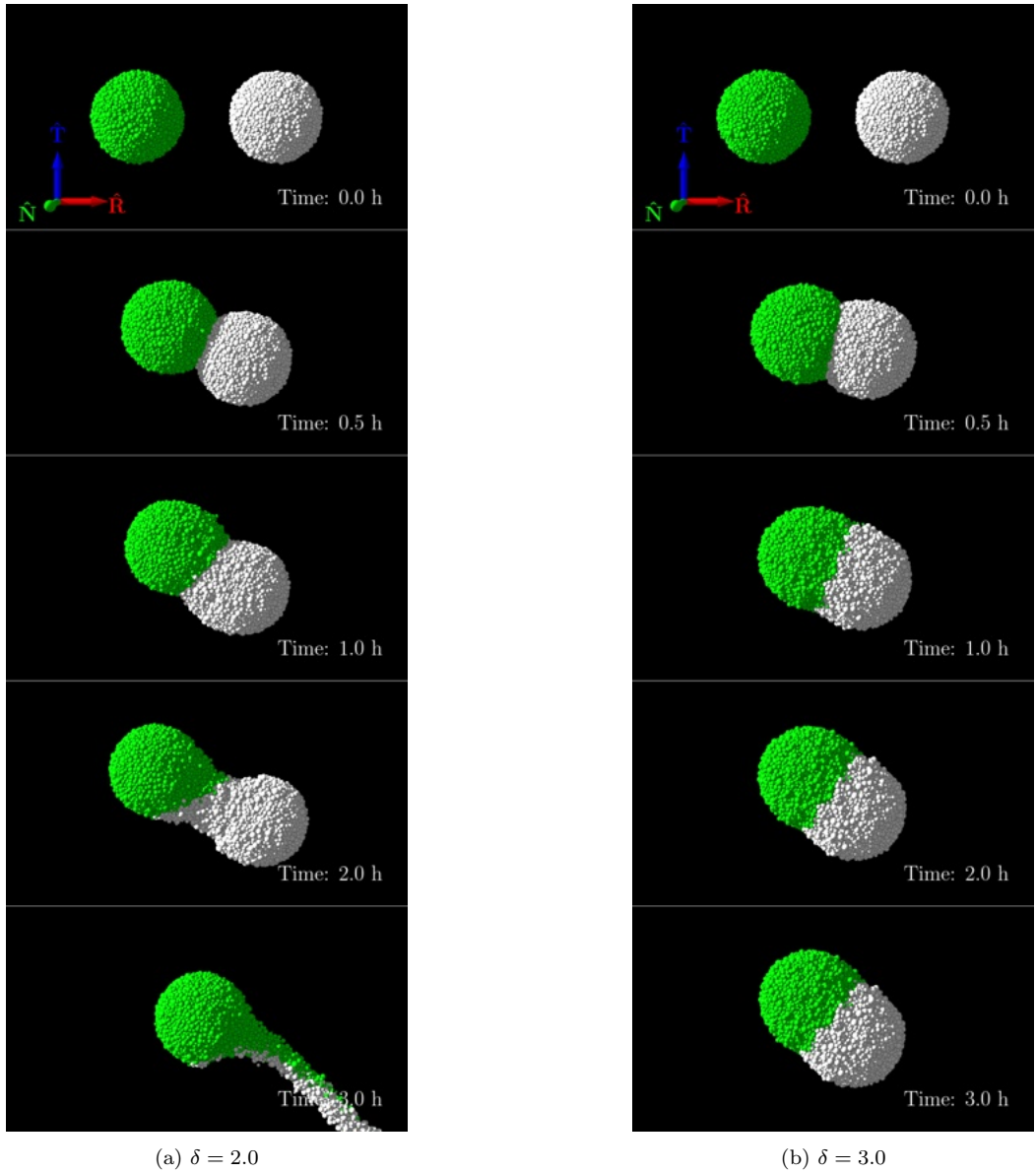

(a) δ = 2.0

(b) δ = 3.0

**Figure 3.** Snapshots of equal-mass ratio collisions with an impact speed of $1v_{\rm esc}$ and a radial (-R) impact geometry for $\delta = 2$ and $\delta = 3$. When the bodies are close-in, the tides from the primary prevent them from fully merging, and instead, they temporarily form a contact binary which is then stripped, resulting in a hit-and-run. Animations of these two simulations can be found in the accompanying Zenodo repository at doi:10.5281/zenodo.15790522.

### 4.4. Effect of Impact Geometry

Here, we keep $M_{\rm proj}/M_{\rm targ}$ fixed to 0.1 and compare the results between the -R and +T impact geometries and the case with no tides. In Figure 8, we show snapshots from three simulations with an impact speed of $v_{\rm imp}/v_{\rm esc} = 10$ at an orbital distance of $\delta = 3.0$ with two different impact geometries and a control case with no tides. In Figure 8(a), tides are ignored although the target has the same pre-impact rotation as if it were synchronously orbiting at $\delta = 3$, while Figures 8(b) and (c) show the -R and +T impact geometries. Despite all three collisions having the same specific impact energy ($Q_R \sim 320$ J kg$^{-1}$), they result in slightly different largest remnant masses, which are driven by the effect of tides. We find that collisions with the +T geometry are much more destructive than the equivalent collisions with the -R geometry. Here, the -R collision results in a largest remnant mass of $M_{\rm lr}/M_{\rm tot} \sim 0.67$, while the +T collision results in a smaller remnant of $M_{\rm lr}/M_{\rm tot} \sim 0.4$. The control case, which includes rotation but not tides, has the largest remnant as expected, with $M_{\rm lr}/M_{\rm tot} \sim 0.69$. This is consistent with the findings of R. Hyodo & K. Ohtsuki (2014), who found that impacts along the tangential direction tend to be the most destructive. Collisions in the azimuthal direction (+T) are typically the most destructive, as ejecta follows the tidal force direction (radially), hindering reaccretion, and the largest remnant becomes radially elongated, making it more prone to tidal stripping.

In Figure 9, we show how the catastrophic disruption threshold for these three cases (no tides, -R, and +T) depends on the orbital distance $\delta$. In this figure, each point is the result of a nonlinear least squares of impact outcomes $(M_{\rm lr}/M_{\rm tot})$ to the specific impact energy $(Q_R)$ to determine the value of $Q_{\rm RD}^{\star}$ that results in the best fit to the universal law (Equation (2)), while the error bars are simply the $1\sigma$ standard deviations from the fit to give a rough sense of its quality. In the case of no tides, $Q_{RD}^\star$ remains fairly constant as a function of the spin rate. This is expected, because previous work by R.-L. Ballouz et al. (2014) demonstrated that target rotation does not significantly affect the impact outcome unless the target is very close to its critical spin limit. We see that $Q_{RD}^\star$ for both the tangential and radial impact geometries decreases abruptly inside of $\delta \sim 3$, which is to be expected, as the strength of tides is increasing. Generally, the radial collisions tend to have a slightly higher $Q_{RD}^\star$ (i.e., less destructive), which is to be expected as well (R. Hyodo & K. Ohtsuki 2014). It is important to note that the value of $Q_{RD}^\star$ for collisions with $\delta \lesssim 2$ are not shown for the -R and +T simulations because the quality of the fits are so poor that the number is effectively meaningless. This is a result of nearly all the simulations resulting in $M_{lr}/M_{tot} \sim 0$.

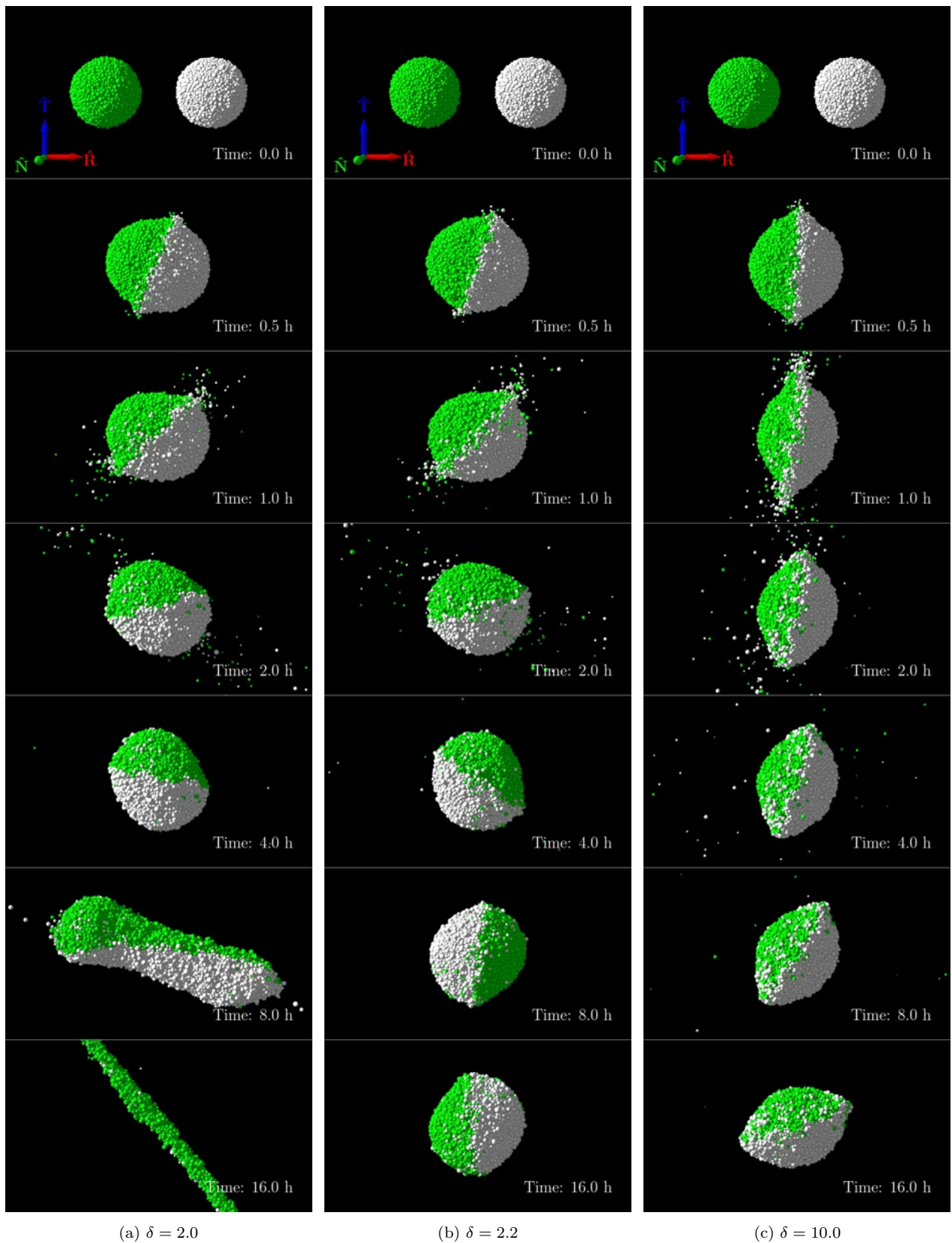

**Figure 4.** Snapshots of equal-mass ratio collisions for different distances with an impact speed of $2v_{\rm esc}$ and a radial (-R) impact geometry. All three collisions have the same specific energy of $Q_R \approx 38.7$ J kg$^{-1}$ 0 yet result in entirely different outcomes as a result of the impact triggering a tidal disruption for close-in targets.

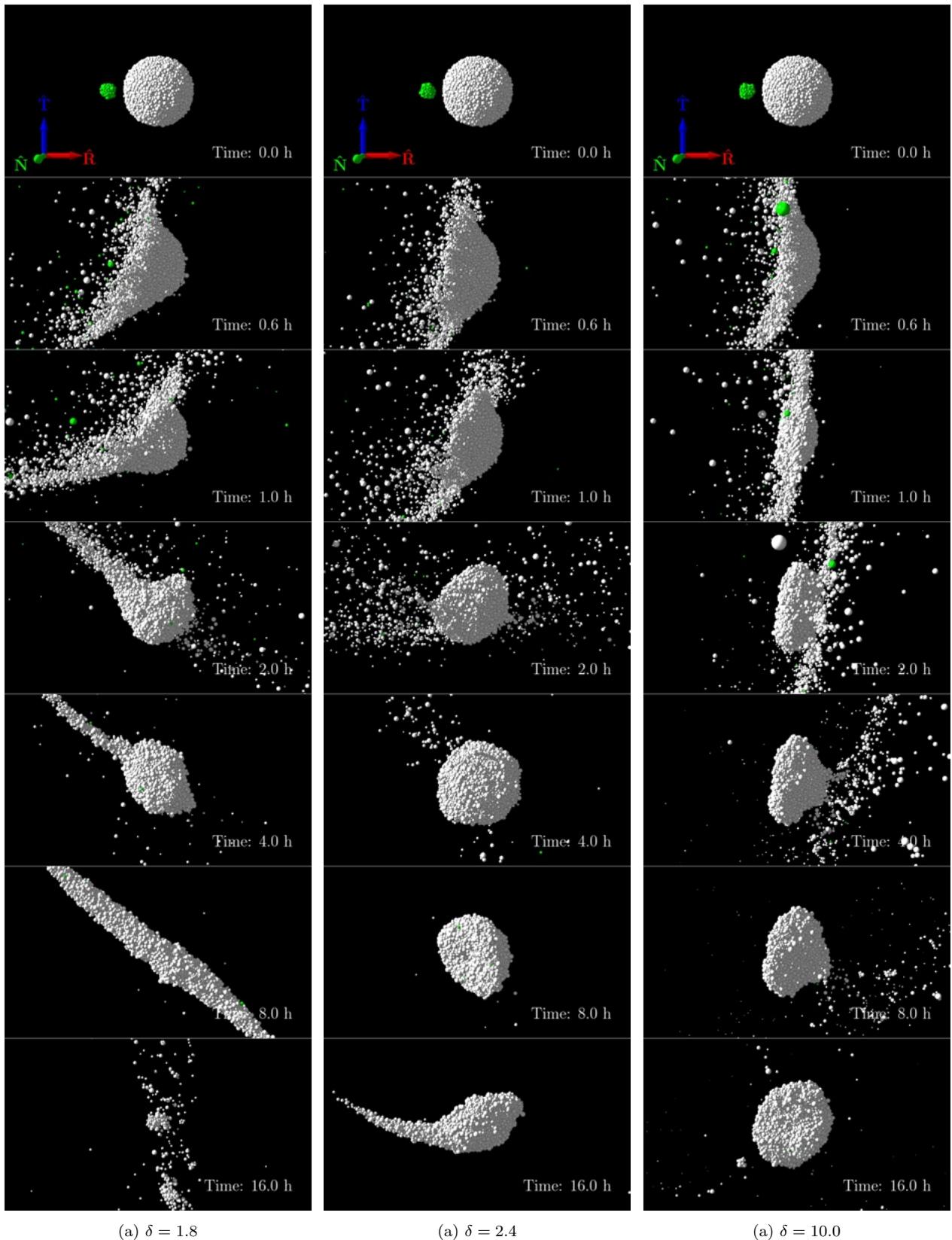

(a) $\delta = 1.8$  (a) $\delta = 2.4$  (a) $\delta = 10.0$

**Figure 5.** Snapshots of collisions with $M_{\mathrm{proj}}/M_{\mathrm{targ}} = 0.01$ at different distances with an impact speed of $30v_{\mathrm{esc}}$ and the radial (-R) impact geometry. All three collisions have the same specific energy of $Q_R \approx 352$ J kg$^{-1}$. Animations of these three simulations can be found in the accompanying Zenodo repository at doi:10.5281/zenodo.15790522.

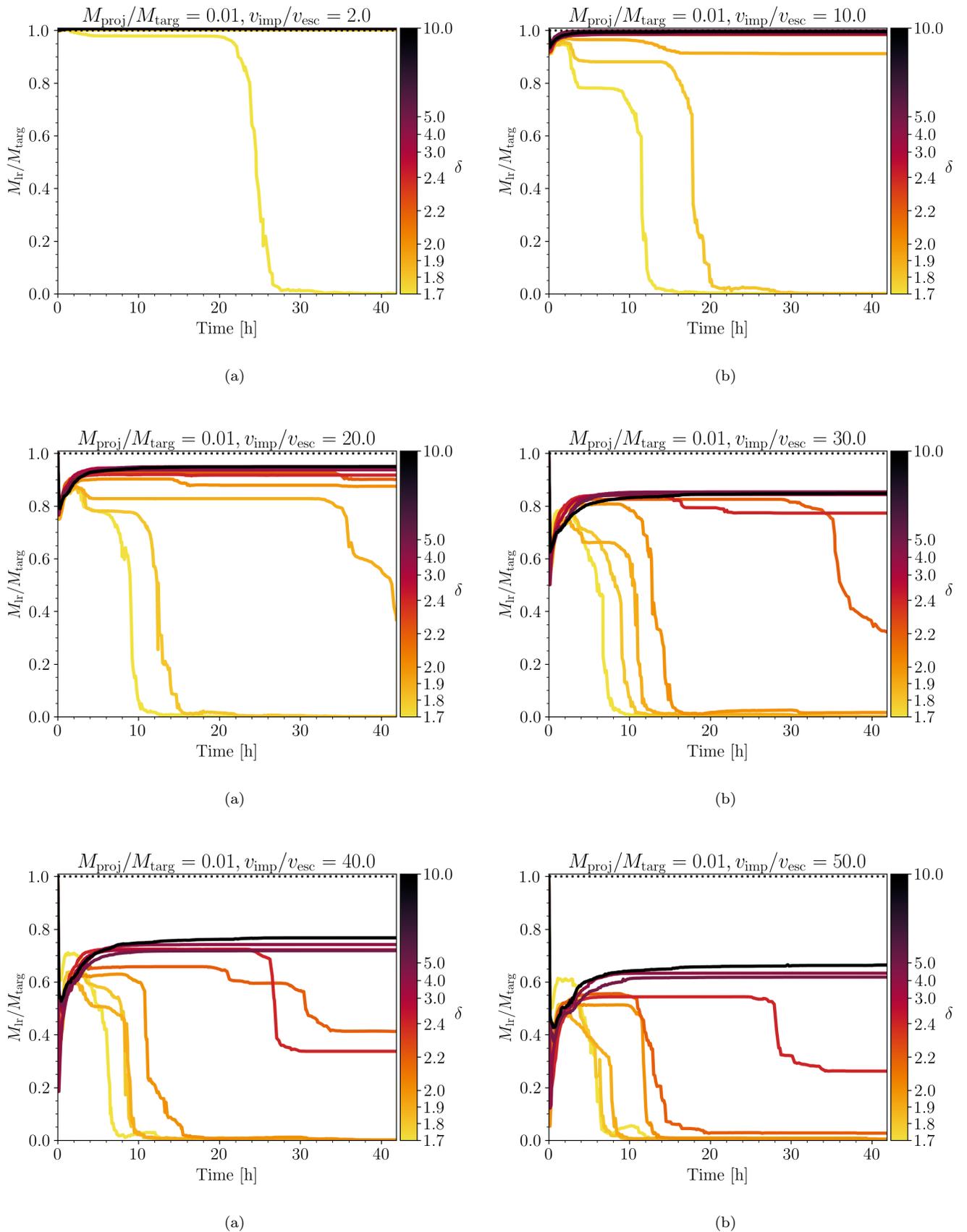

**Figure 6.** The same as Figure 2, but for $M_{\rm proj}/M_{\rm targ} = 0.01$ impacts with the -R impact geometry.

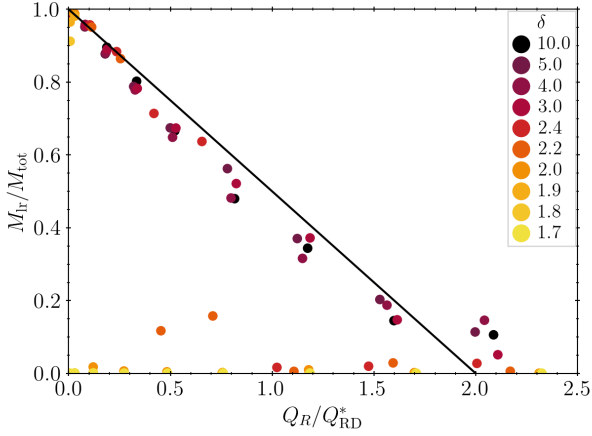

**Figure 7.** Mass fraction of the largest remnant as a function of $Q_R/Q_{RD}^\star$, for orbital distance $\delta$ considering only radial (-R) collisions with $M_{proj}/M_{targ} = 0.1$. A unique $Q_{RD}^\star$ is determined for each fixed $\delta$. For each set of dots corresponding to a fixed $\delta$, a unique $Q_{RD}^\star$ is determined by performing a linear fit to the universal disruption law (Equation (2)). The simulations match the "universal law" quite well for $\delta \gtrsim 3$

### 4.5. Effect of Orbital Distance

As a function of normalized orbital distance, $\delta$, we plot the value of $Q_{RD}^\star$ along with its $1\sigma$ uncertainty that was fit to each set of simulations with the given mass ratio. This plot only shows results with the -R impact geometry, because this is the impact geometry in which all three mass ratios were tested. For higher mass ratios ($M_{proj}/M_{targ} \gtrsim 0.1$) and close orbital distances ($\delta \lesssim 2$), we exclude the derived value for $Q_{RD}^\star$, because each simulation results in a completely disrupted target, even at the lowest impact velocities. This is because we are so close to the target's Roche limit that a large impactor, even at extremely small impact speeds, is sufficiently energetic to disrupt the target. Effectively, this makes the fits for $Q_{RD}^\star$ meaningless, because the mass of the largest remnant in each simulation is almost zero, so it is impossible to fit a linear trend.

It is clear that $Q_{RD}^\star$ decreases with decreasing orbital distance, as would be expected. This trend is most obvious when $M_{proj}/M_{targ} = 0.01$, when the impactor is small enough that it is possible to probe the value for $Q_{RD}^\star$ at small values of $\delta$. At closer distances, the target is spinning faster, and the tidal potential effectively makes the target less gravitationally bound and easier to disrupt.

### 4.6. A Simple Derivation for $Q_{TD}^\star$

Based on the these results, we can derive a crude scaling for how the target's disruption criterion should scale with orbital distance. Z. M. Leinhardt & S. T. Stewart (2012) demonstrated that $Q_{RD}^\star$ is proportional to the target's binding energy per unit mass, with the constant of proportionality depending on the material properties of the target. For a uniform-density sphere of mass $M$ and radius $R$, we can write this as

$$Q_{RD}^\star \propto V_{bind} \propto \frac{3}{5}\frac{GM}{R}, \quad (5)$$

where $G$ is the standard gravitational constant. R.-L. Ballouz et al. (2014) extended this idea to the case of a rotating target. They showed that the rotation-adjusted disruption criterion can be approximated by simply subtracting a rotational potential,

$$Q_{RD,rot}^\star \sim Q_{RD}^\star - V_{rot}. \quad (6)$$

Here, we adopt a similar argument and assume that the tides-adjusted disruption criterion, which we call $Q_{TD}^\star$, should be proportional to the classic disruption criterion, $Q_{RD}^\star$, minus the tidal potential ($V_{tide}$) and the rotational potential ($V_{rot}$): $Q_{TD}^\star \propto Q_{RD}^\star - V_{tide} - V_{rot}$. The tidal potential is of order (C. D. Murray & S. F. Dermott 2000)

$$V_{tide} \sim \frac{GM_P}{d^3}R^2, \quad (7)$$

where $M_P$ is the central body's mass and $d$ is the orbital distance. We can rewrite this in terms of the normalized distance, $\delta$, and obtain

$$V_{tide} \sim \frac{GM}{R}\frac{1}{\delta^3}, \quad (8)$$

where $M$ is the target's mass. We can see that the $GM/R$ term looks like the binding energy of a sphere per unit mass. For the rotational potential, we have

$$V_{rot} \sim \frac{1}{5}R^2\omega^2, \quad (9)$$

where $\omega$ is the target's spin. If we assume the target is in synchronous rotation, we can use the normalized distance and spin ($\delta$ and $\Omega$) to rewrite this in the same form as the tidal potential term:

$$V_{rot} \sim \frac{1}{5}\frac{GM}{R}\frac{1}{\delta^3}. \quad (10)$$

We can see that $V_{rot}$ and $V_{tide}$ look almost identical, with the rotational term being about 5 times weaker. Moreover, both terms have the same scaling with mass and radius as $V_{bind}$, which is proportional to $Q_{RD}^\star$. So we can combine these terms, ignoring constants for now, and arrive at a simple expression for the behavior of $Q_{TD}^\star$:

$$Q_{TD}^\star \sim Q_{RD}^\star\left(1 - \frac{C}{\delta^3}\right), \quad (11)$$

where $C$ is some constant that should capture the effect of both tides and rotation. The scaling of this simple expression makes physical sense: as the orbital distance decreases, tides become stronger and $Q_{TD}^\star$ decreases in proportion. Conversely, for distant satellites, where tides and rotation are both negligible, $Q_{TD}^\star$ converges to $Q_{RD}^\star$. Because $Q_{RD}^\star$ is already determined by fitting Equation (2), we can avoid unnecessary overfitting of a small data set and instead simply select a value for the constant $C$ using physical arguments. When a satellite is at its Roche limit, then its $Q^\star_{\rm TD}$ should effectively go to zero, because at this exact location, any small impact should trigger a catastrophic disruption. In this case, we can simply set $C$ to the cube of the normalized Roche limit so that the term in parentheses goes to zero at the Roche limit. In other words, we can say $C = \delta^3_{\rm Roche}$. Even for a uniform-density ellipsoidal object, the Roche limit is a complicated function of a body's shape and material properties (K. A. Holsapple & P. Michel 2006, 2008). But for a cohesionless, spherical object with a 35° friction angle, such as the objects modeled in these simulations, the dimensionless Roche limit is $\delta_{\rm Roche} \sim 1.5$. So in this case, we have

$$Q^\star_{\rm TD} = Q^\star_{\rm RD}\left(1 - \frac{1.5^3}{\delta^3}\right). \quad (12)$$

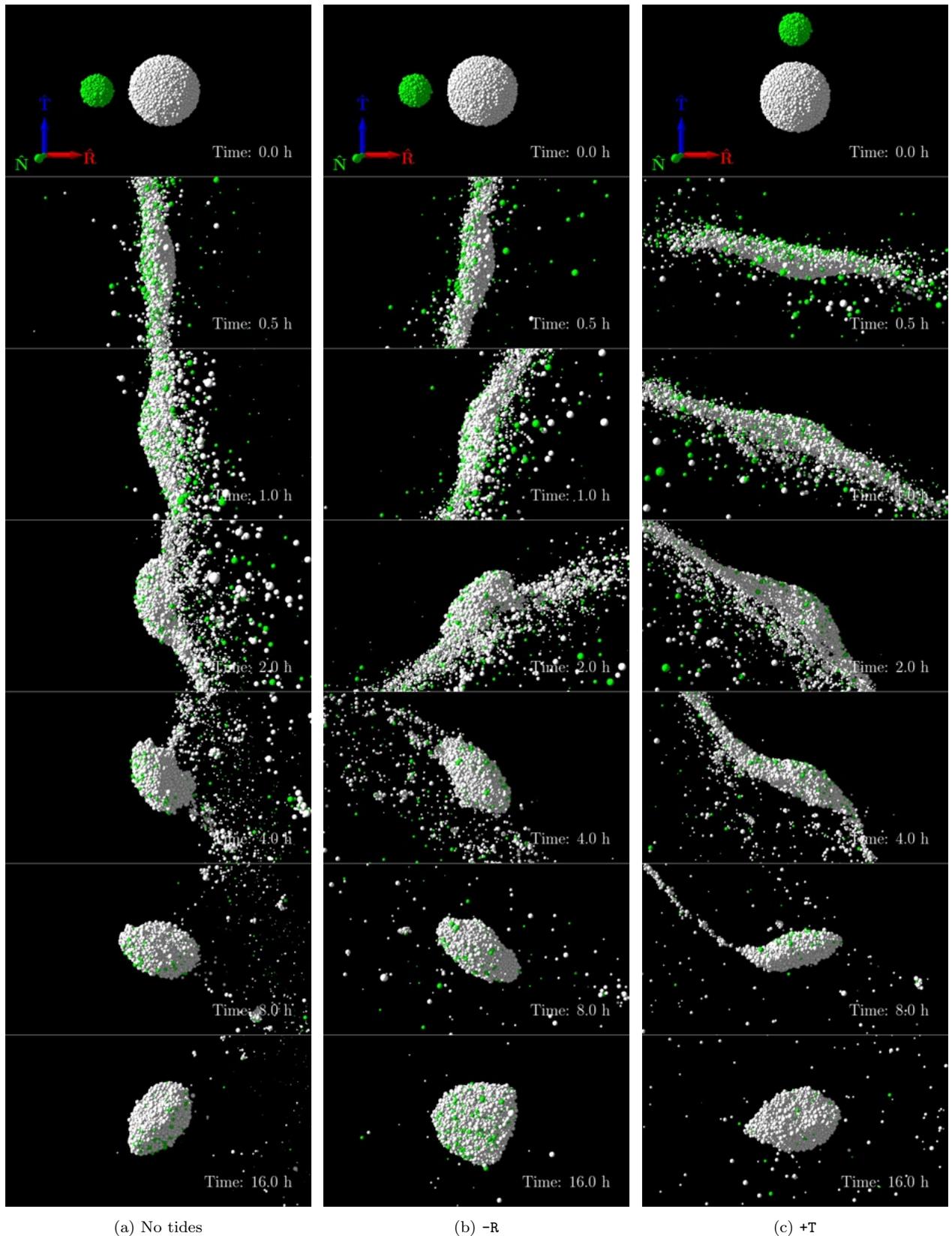

**Figure 8.** Snapshots of simulations with $M_{\rm proj}/M_{\rm targ} = 0.1$, an impact speed of $v_{\rm imp} = 10 v_{\rm esc}$, at an orbital distance of $\delta = 3$ for the control case with no tidal potential, and the two impact geometries: radial (-R) and tangential (+T). For the case with no tides, the target still has the same pre-impact spin rate as if it were in synchronous rotation on an orbit at $\delta = 3$. Animations of these three simulations can be found in the accompanying Zenodo repository at doi:10.5281/zenodo.15790522.

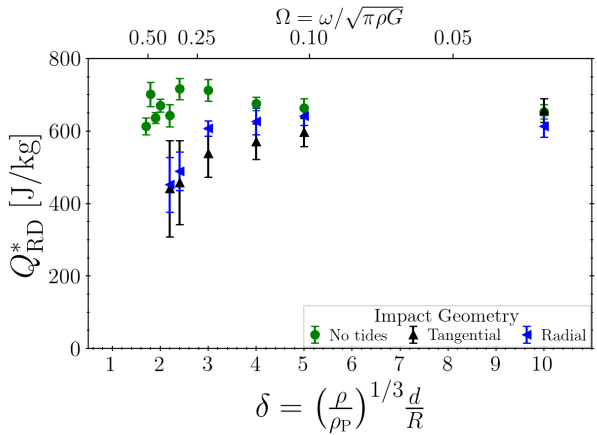

**Figure 9.** Best-fit values of $Q^\star_{\rm RD}$ for simulations with $M_{\rm proj}/M_{\rm targ} = 0.1$ as a function of normalized distance. A control case with no tides (but with rotation) is indicated with green dots, while the tangential (+T) and radial (-R) impact geometries are, respectively, indicated with black and blue triangles. The second x-axis indicates the target's rotation rate. The error bars show the $1\sigma$ standard deviations resulting from nonlinear least squares to give a rough indication of the quality of the fit. The fits get progressively worse with decreasing orbital distance, as a result of the "universal law" breaking down due to the influence of tides. Overall, tangential collisions tend to be slightly more destructive (i.e., a lower $Q^\star_{\rm RD}$).

Note that for objects with different material properties or different shapes, $\delta_{\rm Roche}$ will vary accordingly. In this case, $Q^\star_{\rm RD}$ will also vary, being a function of the target properties as well. Sticking with a simple fixed value of 1.5 is probably appropriate for most situations, given uncertainty in many other parameters.

In Figure 10(a), we show $Q^\star_{\rm RD}$ as a function of normalized orbital distance for each impactor mass ratio for -R collisions. We can clearly see that for $\delta \gtrsim 3$, the catastrophic disruption threshold remains relatively constant. As the distance decreases, $Q^\star_{\rm RD}$ drops off rapidly as might be expected from Equation (12). In Figure 10(b), we show the exactly same plot with the approximation for $Q^\star_{\rm TD}$ shown (Equation (12)). We note that the curves for $Q^\star_{\rm TD}$ are not fit to the simulation results. Rather, the constant $C$ is assumed to be 1.5, and we simply take the value for $Q^\star_{\rm RD}$ at $\delta = 10$ as an approximation for $Q^\star_{\rm RD}$ at an infinitely large distance. It seems that our expression for $Q^\star_{\rm RD}$ does a good job of describing the distance dependence of catastrophic disruption, given its simplicity.

## 5. Discussion

Through simple physical arguments and numerical simulations, we demonstrated that the catastrophic disruption criterion should be a strong function of orbital distance for close-in satellites. Of course, there are many small satellites in the solar system that orbit close to their host planet, so we can ask what these results should imply for these bodies. In Figure 11, we plot the simulation results and scaling for $Q^\star_{\rm TD}$ (identical to Figure 10), with the orbital distances of every close-in satellite in the solar system where possible.[3] We can see immediately that many satellites orbit close to their planet, in a regime where their catastrophic disruption threshold should be dropping off as a result of tides. In terms of $\delta$, the closest moons are Neptune's Naiad, Jupiter's Metis and Adrastea, and Saturn's Daphnis, but other notable moons are located in such a regime, such as Mars's Phobos. In addition, most small binary asteroids have satellites that orbit within just a few primary radii (e.g., P. Pravec et al. 2019). A list in order of increasing $\delta$ of the satellites shown in Figure 10 is tabulated in Table 3 in Appendix B.

Because these various satellites have different formation histories and dynamical environments, and this study is focused more broadly, we cannot say anything in detail about a particular satellite, although we can draw some general conclusions. Many of these small, close-in satellites orbit in a regime where they should be particularly sensitive to catastrophic disruption. Yet, these satellites still clearly exist. Although a good theorist might conclude that something must be wrong with the observations, we will attempt to draw separate conclusions. Broadly speaking, we can make one of three conclusions: (1) many of these satellites must be relatively young because they should get destroyed often by collisions, (2) despite their sensitivity to catastrophic disruption, these satellites are old because collisions are rare, or (3) these satellites have much higher impact strengths than are modeled in this work and are therefore less prone to catastrophic disruption. To elaborate on point #3, these simulations may not completely reflect the reality for most of these satellites, as this study (in order to keep things tractable) considered only cohesionless, spherical rubble piles having a friction angle of 35°. However, it is entirely plausible that some of these satellites have more significant impact strength. In fact, if the age of a satellite can be determined independently, then this scaling law could be used to infer its minimum strength required to survive against impacts over its lifetime.

In reality, the answer could be different for each satellite or satellite system, or some combination of all three: some satellites are weak and young, some are old and strong, or some rarely have large collisions, and their age and strength are poorly correlated. The important takeaway is that impact outcomes on close-in satellites should deviate significantly from the predictions of traditional scaling laws or numerical models. This is an important caveat to keep in mind when studying the evolution of these bodies.

---
[3] Because the normalized orbital distance $\delta$ depends on the satellite's bulk density, we only include satellites for which there is a measured GM and bulk radius, which we combine to determine a density.

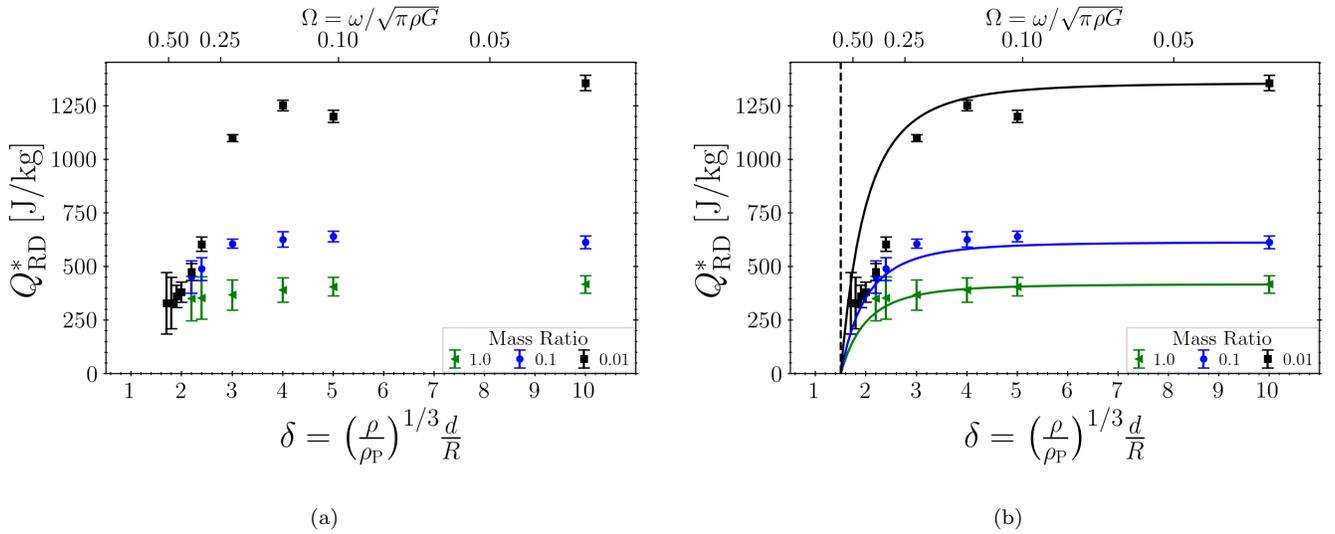

**Figure 10.** (a) Plots of $Q^\star_{\rm RD}$ for collisions for all three mass ratio collisions and radial (-R) impacts. (b) The same plot as (a), with the scaling from Equation (12) to account for tides. Note: this scaling is not fit to the data, but rather the value of $Q^\star_{\rm RD}$ at $\delta = 10$ is taken as a notional value for the value of $Q^\star_{\rm RD}$ at an infinite distance. This simple scaling does a remarkable job of predicting the effective $Q^\star_{\rm RD}$ at close orbital distances, despite its simplicity. The black vertical dashed line indicates the approximate Roche limit for a cohesionless, spherical satellite with a 35° friction angle and demonstrates that tides play an important role in collisional disruptions well outside the Roche limit.

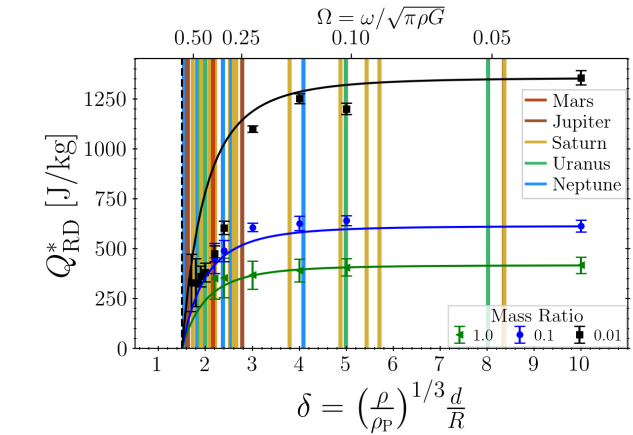

**Figure 11.** The same plot from Figure 10, with the normalized distance $\delta$ for each natural satellite with a measured bulk density around Mars, Jupiter, Saturn, Uranus, and Neptune. Many of the small moonlets around the outer planets orbit in a regime where their catastrophic disruption threshold should be significantly smaller than that derived from scaling laws or numerical models that do not account for tides. The names and orbital distances are listed in Table 3 in Appendix B.

We hope that the simple scaling derived in this work for $Q^\star_{\rm TD}$ will be helpful for future studies about the collisional or dynamical evolution of close-in satellites. For example, $N$-body and other numerical codes that use prescription-based methods to determine collision outcomes according to scaling laws could easily adopt our scaling for $Q^\star_{\rm TD}$. These tools are already designed to compute $Q^\star_{\rm RD}$ according to some scaling law, and with a single line of code, this could be modified to $Q^\star_{\rm TD}$ in cases where a collision takes place close to the central planet.

## 6. Conclusions

Although it is intuitive that the catastrophic disruption threshold should depend on the tides of an external body, this is the first time this effect has been quantified to our knowledge. For close-in satellites, we demonstrated that catastrophic collision outcomes tend to be bimodal, as tides are often capable of fully disrupting a body that was only partially disrupted by the initial collision. There is also some time dependence here, where satellites may take several orbit periods to fully disrupt following an impact.

We showed that the "universal law" for catastrophic disruptions breaks down inside a normalized orbital distance of $\delta \sim 3$, which is consistent with the findings of R. Hyodo & K. Ohtsuki (2014). A simple scaling for the tidal-dependent catastrophic disruption threshold was derived, which shows the disruption threshold is inversely proportional to the cube of the orbital distance. We hope this scaling can easily be adopted by other numerical codes that incorporate prescription-based methods to handle catastrophic collisions. In addition, it can guide studies of internal structures and material properties of small satellites, which need to survive collisional bombardment over their estimated ages.

Many of the small satellites in the solar system have orbits within $\delta \sim 3$, meaning that in principle, they should be much more susceptible to catastrophic disruption than traditional scaling laws that neglect tides would predict. This may be an important caveat when considering their orbital and dynamical evolution. Additionally, many multiasteroid systems have close-in satellites where these scaling laws may be important for determining their collisional lifetimes.

## Acknowledgments

We thank Matija Ćuk for helpful discussions. H.A. was supported by the French government, through the UCA J.E.D.I. Investments in the Future project managed by the National Research Agency (ANR) with the reference number ANR-15-IDEX-01. The simulations were performed on the Mesocentre

SIGAMM machine, hosted by the Observatoire de la Cote d'Azur. We also thank CNES for financial support.

*Software:* PKDRAV (D. C. Richardson et al. 2000; S. R. Schwartz et al. 2014; Y. Zhang et al. 2017), Persistence of Vision Raytracer (https://www.povray.org/), numpy, (C. R. Harris et al. 2020), scipy (P. Virtanen et al. 2020).

# Appendix A
# Simulation Tables

Here, we list all the simulations used in this study. Table 1 lists the main simulation suite, and Table 2 lists the control simulations where no central body is present.

**Table 1**
Table for Full Simulation Suite

| $\frac{M_{proj}}{M_{targ}}$ | $\delta$ | $\frac{v_{imp}}{v_{esc}}$ | $Q_R$ | Dir | $\frac{M_{lr}}{M_{tot}}$ | $Q_{RD}^\star$ |
|---|---|---|---|---|---|---|
| 1.0 | 1.7 | 1.0 | 9.68 | -R | 0.49 | |
| 1.0 | 1.7 | 2.0 | 38.7 | -R | 0.0 | |
| 1.0 | 1.7 | 4.0 | 155 | -R | 0.0 | N/A |
| 1.0 | 1.7 | 6.0 | 348 | -R | 0.0 | |
| 1.0 | 1.7 | 8.0 | 619 | -R | 0.0 | |
| 1.0 | 1.7 | 10.0 | 968 | -R | 0.0 | |
| 1.0 | 1.8 | 1.0 | 9.68 | -R | 0.48 | |
| 1.0 | 1.8 | 2.0 | 38.7 | -R | 0.0 | |
| 1.0 | 1.8 | 4.0 | 155 | -R | 0.0 | N/A |
| 1.0 | 1.8 | 6.0 | 348 | -R | 0.0 | |
| 1.0 | 1.8 | 8.0 | 619 | -R | 0.0 | |
| 1.0 | 1.8 | 10.0 | 968 | -R | 0.0 | |
| 1.0 | 1.9 | 1.0 | 9.68 | -R | 0.47 | |
| 1.0 | 1.9 | 2.0 | 38.7 | -R | 0.01 | |
| 1.0 | 1.9 | 4.0 | 155 | -R | 0.01 | N/A |
| 1.0 | 1.9 | 6.0 | 348 | -R | 0.0 | |
| 1.0 | 1.9 | 8.0 | 619 | -R | 0.0 | |
| 1.0 | 1.9 | 10.0 | 968 | -R | 0.0 | |
| 1.0 | 2.0 | 1.0 | 9.68 | -R | 0.48 | |
| 1.0 | 2.0 | 2.0 | 38.7 | -R | 0.01 | |
| 1.0 | 2.0 | 4.0 | 155 | -R | 0.01 | N/A |
| 1.0 | 2.0 | 6.0 | 348 | -R | 0.0 | |
| 1.0 | 2.0 | 8.0 | 619 | -R | 0.0 | |
| 1.0 | 2.0 | 10.0 | 968 | -R | 0.0 | |
| 1.0 | 2.2 | 1.0 | 9.68 | -R | 0.45 | |
| 1.0 | 2.2 | 2.0 | 38.7 | -R | 1.0 | |
| 1.0 | 2.2 | 4.0 | 155 | -R | 0.01 | 352 |
| 1.0 | 2.2 | 6.0 | 348 | -R | 0.02 | |
| 1.0 | 2.2 | 8.0 | 619 | -R | 0.0 | |
| 1.0 | 2.2 | 10.0 | 968 | -R | 0.0 | |
| 1.0 | 2.4 | 1.0 | 9.68 | -R | 0.48 | |
| 1.0 | 2.4 | 2.0 | 38.7 | -R | 1.0 | |
| 1.0 | 2.4 | 4.0 | 155 | -R | 0.08 | 354 |
| 1.0 | 2.4 | 6.0 | 348 | -R | 0.02 | |
| 1.0 | 2.4 | 8.0 | 619 | -R | 0.01 | |
| 1.0 | 2.4 | 10.0 | 968 | -R | 0.0 | |
| 1.0 | 3.0 | 1.0 | 9.68 | -R | 1.0 | |
| 1.0 | 3.0 | 2.0 | 38.7 | -R | 1.0 | |
| 1.0 | 3.0 | 4.0 | 155 | -R | 0.37 | 369 |
| 1.0 | 3.0 | 6.0 | 348 | -R | 0.08 | |
| 1.0 | 3.0 | 8.0 | 619 | -R | 0.02 | |
| 1.0 | 3.0 | 10.0 | 968 | -R | 0.0 | |
| 1.0 | 4.0 | 1.0 | 9.68 | -R | 1.0 | |
| 1.0 | 4.0 | 2.0 | 38.7 | -R | 1.0 | |
| 1.0 | 4.0 | 4.0 | 155 | -R | 0.87 | 390 |
| 1.0 | 4.0 | 6.0 | 348 | -R | 0.16 | |
| 1.0 | 4.0 | 8.0 | 619 | -R | 0.02 | |
| 1.0 | 4.0 | 10.0 | 968 | -R | 0.01 | |
| 1.0 | 5.0 | 1.0 | 9.68 | -R | 1.0 | |
| 1.0 | 5.0 | 2.0 | 38.7 | -R | 1.0 | |
| 1.0 | 5.0 | 4.0 | 155 | -R | 0.87 | 407 |
| 1.0 | 5.0 | 6.0 | 348 | -R | 0.38 | |
| 1.0 | 5.0 | 8.0 | 619 | -R | 0.02 | |
| 1.0 | 5.0 | 10.0 | 968 | -R | 0.01 | |
| 1.0 | 10.0 | 1.0 | 9.68 | -R | 1.0 | |
| 1.0 | 10.0 | 2.0 | 38.7 | -R | 1.0 | |
| 1.0 | 10.0 | 4.0 | 155 | -R | 0.89 | 417 |
| 1.0 | 10.0 | 6.0 | 348 | -R | 0.46 | |
| 1.0 | 10.0 | 8.0 | 619 | -R | 0.04 | |
| 1.0 | 10.0 | 10.0 | 968 | -R | 0.01 | |
| 0.1 | 1.7 | 1.0 | 3.2 | -R | 0.0 | |
| 0.1 | 1.7 | 2.0 | 12.8 | -R | 0.0 | |
| 0.1 | 1.7 | 4.0 | 51.2 | -R | 0.0 | |
| 0.1 | 1.7 | 6.0 | 115 | -R | 0.0 | N/A |
| 0.1 | 1.7 | 8.0 | 205 | -R | 0.0 | |
| 0.1 | 1.7 | 10.0 | 320 | -R | 0.0 | |
| 0.1 | 1.7 | 12.5 | 500 | -R | 0.0 | |
| 0.1 | 1.7 | 15.0 | 720 | -R | 0.0 | |
| 0.1 | 1.7 | 17.5 | 980 | -R | 0.0 | |
| 0.1 | 1.7 | 20.0 | 1280 | -R | 0.0 | |
| 0.1 | 1.8 | 1.0 | 3.2 | -R | 0.91 | |
| 0.1 | 1.8 | 2.0 | 12.8 | -R | 0.0 | |
| 0.1 | 1.8 | 4.0 | 51.2 | -R | 0.0 | |
| 0.1 | 1.8 | 6.0 | 115 | -R | 0.0 | |
| 0.1 | 1.8 | 8.0 | 205 | -R | 0.0 | N/A |
| 0.1 | 1.8 | 10.0 | 320 | -R | 0.0 | |
| 0.1 | 1.8 | 12.5 | 500 | -R | 0.0 | |
| 0.1 | 1.8 | 15.0 | 720 | -R | 0.0 | |
| 0.1 | 1.8 | 17.5 | 980 | -R | 0.0 | |
| 0.1 | 1.8 | 20.0 | 1280 | -R | 0.0 | |
| 0.1 | 1.9 | 1.0 | 3.2 | -R | 0.97 | |
| 0.1 | 1.9 | 2.0 | 12.8 | -R | 0.98 | |
| 0.1 | 1.9 | 4.0 | 51.2 | -R | 0.0 | |
| 0.1 | 1.9 | 6.0 | 115 | -R | 0.0 | |
| 0.1 | 1.9 | 8.0 | 205 | -R | 0.0 | N/A |
| 0.1 | 1.9 | 10.0 | 320 | -R | 0.0 | |
| 0.1 | 1.9 | 12.5 | 500 | -R | 0.0 | |
| 0.1 | 1.9 | 15.0 | 720 | -R | 0.0 | |
| 0.1 | 1.9 | 17.5 | 980 | -R | 0.0 | |
| 0.1 | 1.9 | 20.0 | 1280 | -R | 0.0 | |
| 0.1 | 2.0 | 1.0 | 3.2 | -R | 1.0 | |
| 0.1 | 2.0 | 2.0 | 12.8 | -R | 0.99 | |
| 0.1 | 2.0 | 4.0 | 51.2 | -R | 0.02 | |
| 0.1 | 2.0 | 6.0 | 115 | -R | 0.01 | |
| 0.1 | 2.0 | 8.0 | 205 | -R | 0.0 | N/A |
| 0.1 | 2.0 | 10.0 | 320 | -R | 0.0 | |
| 0.1 | 2.0 | 12.5 | 500 | -R | 0.01 | |
| 0.1 | 2.0 | 15.0 | 720 | -R | 0.0 | |
| 0.1 | 2.0 | 17.5 | 980 | -R | 0.0 | |
| 0.1 | 2.0 | 20.0 | 1280 | -R | 0.0 | |

**Table 1**
(Continued)

| $\frac{M_{\text{proj}}}{M_{\text{targ}}}$ | $\delta$ | $\frac{v_{\text{imp}}}{v_{\text{esc}}}$ | $Q_R$ | Dir | $\frac{M_{lr}}{M_{\text{tot}}}$ | $Q_{\text{RD}}^{\star}$ |
|---|---|---|---|---|---|---|
| 0.1 | 2.2 | 1.0  | 3.2  | −R | 1.0  |     |
| 0.1 | 2.2 | 2.0  | 12.8 | −R | 0.99 |     |
| 0.1 | 2.2 | 4.0  | 51.2 | −R | 0.95 |     |
| 0.1 | 2.2 | 6.0  | 115  | −R | 0.86 |     |
| 0.1 | 2.2 | 8.0  | 205  | −R | 0.12 | 452 |
| 0.1 | 2.2 | 10.0 | 320  | −R | 0.16 |     |
| 0.1 | 2.2 | 12.5 | 500  | −R | 0.01 |     |
| 0.1 | 2.2 | 15.0 | 720  | −R | 0.03 |     |
| 0.1 | 2.2 | 17.5 | 980  | −R | 0.01 |     |
| 0.1 | 2.2 | 20.0 | 1280 | −R | 0.01 |     |
| 0.1 | 2.4 | 1.0  | 3.2  | −R | 1.0  |     |
| 0.1 | 2.4 | 2.0  | 12.8 | −R | 0.99 |     |
| 0.1 | 2.4 | 4.0  | 51.2 | −R | 0.96 |     |
| 0.1 | 2.4 | 6.0  | 115  | −R | 0.88 |     |
| 0.1 | 2.4 | 8.0  | 205  | −R | 0.71 | 489 |
| 0.1 | 2.4 | 10.0 | 320  | −R | 0.64 |     |
| 0.1 | 2.4 | 12.5 | 500  | −R | 0.02 |     |
| 0.1 | 2.4 | 15.0 | 720  | −R | 0.02 |     |
| 0.1 | 2.4 | 17.5 | 980  | −R | 0.03 |     |
| 0.1 | 2.4 | 20.0 | 1280 | −R | 0.01 |     |
| 0.1 | 3.0 | 1.0  | 3.2  | −R | 1.0  |     |
| 0.1 | 3.0 | 2.0  | 12.8 | −R | 0.99 |     |
| 0.1 | 3.0 | 4.0  | 51.2 | −R | 0.95 |     |
| 0.1 | 3.0 | 6.0  | 115  | −R | 0.89 |     |
| 0.1 | 3.0 | 8.0  | 205  | −R | 0.78 | 607 |
| 0.1 | 3.0 | 10.0 | 320  | −R | 0.67 |     |
| 0.1 | 3.0 | 12.5 | 500  | −R | 0.52 |     |
| 0.1 | 3.0 | 15.0 | 720  | −R | 0.37 |     |
| 0.1 | 3.0 | 17.5 | 980  | −R | 0.15 |     |
| 0.1 | 3.0 | 20.0 | 1280 | −R | 0.05 |     |
| 0.1 | 4.0 | 1.0  | 3.2  | −R | 1.0  |     |
| 0.1 | 4.0 | 2.0  | 12.8 | −R | 0.99 |     |
| 0.1 | 4.0 | 4.0  | 51.2 | −R | 0.96 |     |
| 0.1 | 4.0 | 6.0  | 115  | −R | 0.88 |     |
| 0.1 | 4.0 | 8.0  | 205  | −R | 0.78 | 627 |
| 0.1 | 4.0 | 10.0 | 320  | −R | 0.65 |     |
| 0.1 | 4.0 | 12.5 | 500  | −R | 0.48 |     |
| 0.1 | 4.0 | 15.0 | 720  | −R | 0.32 |     |
| 0.1 | 4.0 | 17.5 | 980  | −R | 0.19 |     |
| 0.1 | 4.0 | 20.0 | 1280 | −R | 0.15 |     |
| 0.1 | 5.0 | 1.0  | 3.2  | −R | 1.0  |     |
| 0.1 | 5.0 | 2.0  | 12.8 | −R | 0.99 |     |
| 0.1 | 5.0 | 4.0  | 51.2 | −R | 0.95 |     |
| 0.1 | 5.0 | 6.0  | 115  | −R | 0.88 |     |
| 0.1 | 5.0 | 8.0  | 205  | −R | 0.79 | 641 |
| 0.1 | 5.0 | 10.0 | 320  | −R | 0.67 |     |
| 0.1 | 5.0 | 12.5 | 500  | −R | 0.56 |     |
| 0.1 | 5.0 | 15.0 | 720  | −R | 0.37 |     |
| 0.1 | 5.0 | 17.5 | 980  | −R | 0.2  |     |
| 0.1 | 5.0 | 20.0 | 1280 | −R | 0.11 |     |
| 0.1 | 10.0 | 1.0  | 3.2  | −R | 1.0  |     |
| 0.1 | 10.0 | 2.0  | 12.8 | −R | 0.99 |     |
| 0.1 | 10.0 | 4.0  | 51.2 | −R | 0.96 |     |
| 0.1 | 10.0 | 6.0  | 115  | −R | 0.9  |     |
| 0.1 | 10.0 | 8.0  | 205  | −R | 0.8  | 614 |
| 0.1 | 10.0 | 10.0 | 320  | −R | 0.67 |     |
| 0.1 | 10.0 | 12.5 | 500  | −R | 0.48 |     |
| 0.1 | 10.0 | 15.0 | 720  | −R | 0.34 |     |
| 0.1 | 10.0 | 17.5 | 980  | −R | 0.15 |     |
| 0.1 | 10.0 | 20.0 | 1280 | −R | 0.11 |     |
| 0.1 | 1.7 | 1.0  | 3.2  | +T | 0.0  |     |
| 0.1 | 1.7 | 2.0  | 12.8 | +T | 0.0  |     |
| 0.1 | 1.7 | 4.0  | 51.2 | +T | 0.0  |     |
| 0.1 | 1.7 | 6.0  | 115  | +T | 0.0  | N/A |
| 0.1 | 1.7 | 8.0  | 205  | +T | 0.0  |     |
| 0.1 | 1.7 | 10.0 | 320  | +T | 0.0  |     |
| 0.1 | 1.7 | 15.0 | 720  | +T | 0.0  |     |
| 0.1 | 1.7 | 20.0 | 1280 | +T | 0.0  |     |
| 0.1 | 1.8 | 1.0  | 3.2  | +T | 0.0  |     |
| 0.1 | 1.8 | 2.0  | 12.8 | +T | 0.0  |     |
| 0.1 | 1.8 | 4.0  | 51.2 | +T | 0.0  |     |
| 0.1 | 1.8 | 6.0  | 115  | +T | 0.0  | N/A |
| 0.1 | 1.8 | 8.0  | 205  | +T | 0.0  |     |
| 0.1 | 1.8 | 10.0 | 320  | +T | 0.0  |     |
| 0.1 | 1.8 | 15.0 | 720  | +T | 0.0  |     |
| 0.1 | 1.8 | 20.0 | 1280 | +T | 0.0  |     |
| 0.1 | 1.9 | 1.0  | 3.2  | +T | 0.97 |     |
| 0.1 | 1.9 | 2.0  | 12.8 | +T | 0.0  |     |
| 0.1 | 1.9 | 4.0  | 51.2 | +T | 0.0  |     |
| 0.1 | 1.9 | 6.0  | 115  | +T | 0.0  | N/A |
| 0.1 | 1.9 | 8.0  | 205  | +T | 0.0  |     |
| 0.1 | 1.9 | 10.0 | 320  | +T | 0.0  |     |
| 0.1 | 1.9 | 15.0 | 720  | +T | 0.0  |     |
| 0.1 | 1.9 | 20.0 | 1280 | +T | 0.0  |     |
| 0.1 | 2.0 | 1.0  | 3.2  | +T | 0.99 |     |
| 0.1 | 2.0 | 2.0  | 12.8 | +T | 0.96 |     |
| 0.1 | 2.0 | 4.0  | 51.2 | +T | 0.09 |     |
| 0.1 | 2.0 | 6.0  | 115  | +T | 0.0  | N/A |
| 0.1 | 2.0 | 8.0  | 205  | +T | 0.01 |     |
| 0.1 | 2.0 | 10.0 | 320  | +T | 0.0  |     |
| 0.1 | 2.0 | 15.0 | 720  | +T | 0.0  |     |
| 0.1 | 2.0 | 20.0 | 1280 | +T | 0.0  |     |
| 0.1 | 2.2 | 1.0  | 3.2  | +T | 0.99 |     |
| 0.1 | 2.2 | 2.0  | 12.8 | +T | 0.97 |     |
| 0.1 | 2.2 | 4.0  | 51.2 | +T | 0.77 |     |
| 0.1 | 2.2 | 6.0  | 115  | +T | 0.02 |     |
| 0.1 | 2.2 | 8.0  | 205  | +T | 0.02 | 441 |
| 0.1 | 2.2 | 10.0 | 320  | +T | 0.04 |     |
| 0.1 | 2.2 | 15.0 | 720  | +T | 0.0  |     |
| 0.1 | 2.2 | 20.0 | 1280 | +T | 0.0  |     |
| 0.1 | 2.4 | 1.0  | 3.2  | +T | 0.99 |     |
| 0.1 | 2.4 | 2.0  | 12.8 | +T | 0.98 |     |
| 0.1 | 2.4 | 4.0  | 51.2 | +T | 0.89 |     |
| 0.1 | 2.4 | 6.0  | 115  | +T | 0.46 | 458 |
| 0.1 | 2.4 | 8.0  | 205  | +T | 0.04 |     |
| 0.1 | 2.4 | 10.0 | 320  | +T | 0.1  |     |
| 0.1 | 2.4 | 15.0 | 720  | +T | 0.03 |     |
| 0.1 | 2.4 | 20.0 | 1280 | +T | 0.01 |     |
| 0.1 | 3.0 | 1.0  | 3.2  | +T | 1.0  |     |
| 0.1 | 3.0 | 2.0  | 12.8 | +T | 0.98 |     |
| 0.1 | 3.0 | 4.0  | 51.2 | +T | 0.93 |     |
| 0.1 | 3.0 | 6.0  | 115  | +T | 0.82 | 539 |
| 0.1 | 3.0 | 8.0  | 205  | +T | 0.66 |     |
| 0.1 | 3.0 | 10.0 | 320  | +T | 0.4  |     |
| 0.1 | 3.0 | 15.0 | 720  | +T | 0.13 |     |

**Table 1**
(Continued)

| $\frac{M_{proj}}{M_{targ}}$ | $\delta$ | $\frac{v_{imp}}{v_{esc}}$ | $Q_R$ | Dir | $\frac{M_{lr}}{M_{tot}}$ | $Q_{RD}^\star$ |
|---|---|---|---|---|---|---|
| 0.1 | 3.0 | 20.0 | 1280 | +T | 0.03 | |
| 0.1 | 4.0 | 1.0 | 3.2 | +T | 1.0 | |
| 0.1 | 4.0 | 2.0 | 12.8 | +T | 0.99 | |
| 0.1 | 4.0 | 4.0 | 51.2 | +T | 0.95 | |
| 0.1 | 4.0 | 6.0 | 115 | +T | 0.87 | 571 |
| 0.1 | 4.0 | 8.0 | 205 | +T | 0.75 | |
| 0.1 | 4.0 | 10.0 | 320 | +T | 0.58 | |
| 0.1 | 4.0 | 15.0 | 720 | +T | 0.17 | |
| 0.1 | 4.0 | 20.0 | 1280 | +T | 0.04 | |
| 0.1 | 5.0 | 1.0 | 3.2 | +T | 1.0 | |
| 0.1 | 5.0 | 2.0 | 12.8 | +T | 0.98 | |
| 0.1 | 5.0 | 4.0 | 51.2 | +T | 0.94 | |
| 0.1 | 5.0 | 6.0 | 115 | +T | 0.86 | 596 |
| 0.1 | 5.0 | 8.0 | 205 | +T | 0.75 | |
| 0.1 | 5.0 | 10.0 | 320 | +T | 0.62 | |
| 0.1 | 5.0 | 15.0 | 720 | +T | 0.27 | |
| 0.1 | 5.0 | 20.0 | 1280 | +T | 0.04 | |
| 0.1 | 10.0 | 1.0 | 3.2 | +T | 1.0 | |
| 0.1 | 10.0 | 2.0 | 12.8 | +T | 0.99 | |
| 0.1 | 10.0 | 4.0 | 51.2 | +T | 0.95 | |
| 0.1 | 10.0 | 6.0 | 115 | +T | 0.89 | 657 |
| 0.1 | 10.0 | 8.0 | 205 | +T | 0.8 | |
| 0.1 | 10.0 | 10.0 | 320 | +T | 0.69 | |
| 0.1 | 10.0 | 15.0 | 720 | +T | 0.36 | |
| 0.1 | 10.0 | 20.0 | 1280 | +T | 0.11 | |
| 0.01 | 1.7 | 1.0 | 0.391 | −R | 0.98 | |
| 0.01 | 1.7 | 2.0 | 1.56 | −R | 0.0 | |
| 0.01 | 1.7 | 4.0 | 6.25 | −R | 0.0 | |
| 0.01 | 1.7 | 6.0 | 14.1 | −R | 0.0 | |
| 0.01 | 1.7 | 8.0 | 25 | −R | 0.0 | |
| 0.01 | 1.7 | 10.0 | 39.1 | −R | 0.0 | 329 |
| 0.01 | 1.7 | 15.0 | 87.9 | −R | 0.0 | |
| 0.01 | 1.7 | 20.0 | 156 | −R | 0.0 | |
| 0.01 | 1.7 | 30.0 | 352 | −R | 0.0 | |
| 0.01 | 1.7 | 40.0 | 625 | −R | 0.0 | |
| 0.01 | 1.7 | 50.0 | 977 | −R | 0.0 | |
| 0.01 | 1.8 | 1.0 | 0.391 | −R | 0.99 | |
| 0.01 | 1.8 | 2.0 | 1.56 | −R | 0.99 | |
| 0.01 | 1.8 | 4.0 | 6.25 | −R | 0.98 | |
| 0.01 | 1.8 | 6.0 | 14.1 | −R | 0.01 | |
| 0.01 | 1.8 | 8.0 | 25 | −R | 0.0 | |
| 0.01 | 1.8 | 10.0 | 39.1 | −R | 0.0 | 330 |
| 0.01 | 1.8 | 15.0 | 87.9 | −R | 0.0 | |
| 0.01 | 1.8 | 20.0 | 156 | −R | 0.0 | |
| 0.01 | 1.8 | 30.0 | 352 | −R | 0.0 | |
| 0.01 | 1.8 | 40.0 | 625 | −R | 0.0 | |
| 0.01 | 1.8 | 50.0 | 977 | −R | 0.0 | |
| 0.01 | 1.9 | 1.0 | 0.391 | −R | 1.0 | |
| 0.01 | 1.9 | 2.0 | 1.56 | −R | 1.0 | |
| 0.01 | 1.9 | 4.0 | 6.25 | −R | 0.99 | |
| 0.01 | 1.9 | 6.0 | 14.1 | −R | 0.98 | |
| 0.01 | 1.9 | 8.0 | 25 | −R | 0.97 | |
| 0.01 | 1.9 | 10.0 | 39.1 | −R | 0.9 | 362 |
| 0.01 | 1.9 | 15.0 | 87.9 | −R | 0.79 | |
| 0.01 | 1.9 | 20.0 | 156 | −R | 0.37 | |
| 0.01 | 1.9 | 30.0 | 352 | −R | 0.0 | |
| 0.01 | 1.9 | 40.0 | 625 | −R | 0.0 | |
| 0.01 | 1.9 | 50.0 | 977 | −R | 0.0 | |
| 0.01 | 2.0 | 1.0 | 0.391 | −R | 1.0 | |
| 0.01 | 2.0 | 2.0 | 1.56 | −R | 1.0 | |
| 0.01 | 2.0 | 4.0 | 6.25 | −R | 1.0 | 381 |
| 0.01 | 2.0 | 6.0 | 14.1 | −R | 0.99 | |
| 0.01 | 2.0 | 8.0 | 25 | −R | 0.98 | |
| 0.01 | 2.0 | 10.0 | 39.1 | −R | 0.97 | |
| 0.01 | 2.0 | 15.0 | 87.9 | −R | 0.9 | |
| 0.01 | 2.0 | 20.0 | 156 | −R | 0.87 | |
| 0.01 | 2.0 | 30.0 | 352 | −R | 0.02 | |
| 0.01 | 2.0 | 40.0 | 625 | −R | 0.0 | |
| 0.01 | 2.0 | 50.0 | 977 | −R | 0.01 | |
| 0.01 | 2.2 | 1.0 | 0.391 | −R | 1.0 | |
| 0.01 | 2.2 | 2.0 | 1.56 | −R | 1.0 | |
| 0.01 | 2.2 | 4.0 | 6.25 | −R | 0.99 | |
| 0.01 | 2.2 | 6.0 | 14.1 | −R | 0.99 | |
| 0.01 | 2.2 | 8.0 | 25 | −R | 0.99 | |
| 0.01 | 2.2 | 10.0 | 39.1 | −R | 0.98 | 475 |
| 0.01 | 2.2 | 15.0 | 87.9 | −R | 0.96 | |
| 0.01 | 2.2 | 20.0 | 156 | −R | 0.89 | |
| 0.01 | 2.2 | 30.0 | 352 | −R | 0.32 | |
| 0.01 | 2.2 | 40.0 | 625 | −R | 0.41 | |
| 0.01 | 2.2 | 50.0 | 977 | −R | 0.03 | |
| 0.01 | 2.4 | 1.0 | 0.391 | −R | 1.0 | |
| 0.01 | 2.4 | 2.0 | 1.56 | −R | 1.0 | |
| 0.01 | 2.4 | 4.0 | 6.25 | −R | 0.99 | |
| 0.01 | 2.4 | 6.0 | 14.1 | −R | 0.99 | |
| 0.01 | 2.4 | 8.0 | 25 | −R | 0.98 | |
| 0.01 | 2.4 | 10.0 | 39.1 | −R | 0.98 | 605 |
| 0.01 | 2.4 | 15.0 | 87.9 | −R | 0.94 | |
| 0.01 | 2.4 | 20.0 | 156 | −R | 0.91 | |
| 0.01 | 2.4 | 30.0 | 352 | −R | 0.77 | |
| 0.01 | 2.4 | 40.0 | 625 | −R | 0.33 | |
| 0.01 | 2.4 | 50.0 | 977 | −R | 0.26 | |
| 0.01 | 3.0 | 1.0 | 0.391 | −R | 1.0 | |
| 0.01 | 3.0 | 2.0 | 1.56 | −R | 1.0 | |
| 0.01 | 3.0 | 4.0 | 6.25 | −R | 1.0 | |
| 0.01 | 3.0 | 6.0 | 14.1 | −R | 0.99 | |
| 0.01 | 3.0 | 8.0 | 25 | −R | 0.99 | |
| 0.01 | 3.0 | 10.0 | 39.1 | −R | 0.99 | 1100 |
| 0.01 | 3.0 | 15.0 | 87.9 | −R | 0.97 | |
| 0.01 | 3.0 | 20.0 | 156 | −R | 0.94 | |
| 0.01 | 3.0 | 30.0 | 352 | −R | 0.84 | |
| 0.01 | 3.0 | 40.0 | 625 | −R | 0.71 | |
| 0.01 | 4.0 | 1.0 | 0.391 | −R | 1.0 | |
| 0.01 | 4.0 | 2.0 | 1.56 | −R | 1.0 | |
| 0.01 | 4.0 | 4.0 | 6.25 | −R | 0.99 | |
| 0.01 | 4.0 | 6.0 | 14.1 | −R | 0.99 | |
| 0.01 | 4.0 | 8.0 | 25 | −R | 0.99 | |
| 0.01 | 4.0 | 10.0 | 39.1 | −R | 0.98 | 1250 |
| 0.01 | 4.0 | 15.0 | 87.9 | −R | 0.96 | |
| 0.01 | 4.0 | 20.0 | 156 | −R | 0.93 | |
| 0.01 | 4.0 | 30.0 | 352 | −R | 0.85 | |
| 0.01 | 4.0 | 40.0 | 625 | −R | 0.73 | |
| 0.01 | 4.0 | 50.0 | 977 | −R | 0.63 | |
| 0.01 | 5.0 | 1.0 | 0.391 | −R | 1.0 | |
| 0.01 | 5.0 | 2.0 | 1.56 | −R | 1.0 | |
| 0.01 | 5.0 | 4.0 | 6.25 | −R | 0.99 | |
| 0.01 | 5.0 | 6.0 | 14.1 | −R | 0.99 | |
| 0.01 | 5.0 | 8.0 | 25 | −R | 0.99 | |
| 0.01 | 5.0 | 10.0 | 39.1 | −R | 0.99 | 1200 |
| 0.01 | 5.0 | 15.0 | 87.9 | −R | 0.97 | |
| 0.01 | 5.0 | 20.0 | 156 | −R | 0.94 | |
| 0.01 | 5.0 | 30.0 | 352 | −R | 0.84 | |

**Table 1**
(Continued)

| $\frac{M_{proj}}{M_{targ}}$ | $\delta$ | $\frac{v_{imp}}{v_{esc}}$ | $Q_R$ | Dir | $\frac{M_{lr}}{M_{tot}}$ | $Q_{RD}^\star$ |
|---|---|---|---|---|---|---|
| 0.01 | 5.0 | 40.0 | 625 | -R | 0.71 | |
| 0.01 | 5.0 | 50.0 | 977 | -R | 0.61 | |
| 0.01 | 10.0 | 1.0 | 0.391 | -R | 1.0 | |
| 0.01 | 10.0 | 2.0 | 1.56 | -R | 1.0 | |
| 0.01 | 10.0 | 4.0 | 6.25 | -R | 1.0 | |
| 0.01 | 10.0 | 6.0 | 14.1 | -R | 0.99 | |
| 0.01 | 10.0 | 8.0 | 25 | -R | 0.99 | |
| 0.01 | 10.0 | 10.0 | 39.1 | -R | 0.99 | 1360 |
| 0.01 | 10.0 | 15.0 | 87.9 | -R | 0.96 | |
| 0.01 | 10.0 | 20.0 | 156 | -R | 0.94 | |
| 0.01 | 10.0 | 30.0 | 352 | -R | 0.84 | |
| 0.01 | 10.0 | 40.0 | 625 | -R | 0.76 | |
| 0.01 | 10.0 | 50.0 | 977 | -R | 0.66 | |

**Note.** From left to right, each column lists the simulation's projectile-to-target mass ratio ($M_{proj}/M_{targ}$), the normalized orbital distance ($\delta$), the impact speed normalized to the target escape speed ($v_{imp}/v_{esc}$), specific impact energy in J kg$^{-1}$ ($Q_R$), impactor direction (Dir), and then the resulting largest remnant mass ($M_{lr}/M_{tot}$) and the best-fit value for the catastrophic disruption threshold, in J kg$^{-1}$ ($Q_{RD}^\star$). Simulations where a meaningful fit could not be determined have N/A listed in the $Q_{RD}^\star$ column.

**Table 2**
Table for Control Runs, where Tides Are Not Included

| $\frac{M_{proj}}{M_{targ}}$ | $\delta$ | $\frac{v_{imp}}{v_{esc}}$ | $Q_R$ | $\Omega$ | $\frac{M_{lr}}{M_{tot}}$ | $Q_{RD}^\star$ |
|---|---|---|---|---|---|---|
| 0.1 | 1.7 | 1.0 | 3.2 | 0.521 | 1.0 | |
| 0.1 | 1.7 | 2.0 | 12.8 | 0.521 | 0.99 | |
| 0.1 | 1.7 | 4.0 | 51.2 | 0.521 | 0.95 | |
| 0.1 | 1.7 | 6.0 | 115 | 0.521 | 0.88 | |
| 0.1 | 1.7 | 8.0 | 205 | 0.521 | 0.79 | 613 |
| 0.1 | 1.7 | 10.0 | 320 | 0.521 | 0.69 | |
| 0.1 | 1.7 | 12.5 | 500 | 0.521 | 0.51 | |
| 0.1 | 1.7 | 15.0 | 720 | 0.521 | 0.33 | |
| 0.1 | 1.7 | 17.5 | 980 | 0.521 | 0.19 | |
| 0.1 | 1.7 | 20.0 | 1280 | 0.521 | 0.06 | |
| 0.1 | 1.8 | 1.0 | 3.2 | 0.478 | 1.0 | |
| 0.1 | 1.8 | 2.0 | 12.8 | 0.478 | 0.99 | |
| 0.1 | 1.8 | 4.0 | 51.2 | 0.478 | 0.94 | |
| 0.1 | 1.8 | 6.0 | 115 | 0.478 | 0.87 | |
| 0.1 | 1.8 | 8.0 | 205 | 0.478 | 0.79 | 701 |
| 0.1 | 1.8 | 10.0 | 320 | 0.478 | 0.69 | |
| 0.1 | 1.8 | 12.5 | 500 | 0.478 | 0.56 | |
| 0.1 | 1.8 | 15.0 | 720 | 0.478 | 0.41 | |
| 0.1 | 1.8 | 17.5 | 980 | 0.478 | 0.3 | |
| 0.1 | 1.8 | 20.0 | 1280 | 0.478 | 0.2 | |
| 0.1 | 1.9 | 1.0 | 3.2 | 0.441 | 1.0 | |
| 0.1 | 1.9 | 2.0 | 12.8 | 0.441 | 0.99 | |
| 0.1 | 1.9 | 4.0 | 51.2 | 0.441 | 0.96 | |
| 0.1 | 1.9 | 6.0 | 115 | 0.441 | 0.9 | |
| 0.1 | 1.9 | 8.0 | 205 | 0.441 | 0.8 | 636 |
| 0.1 | 1.9 | 10.0 | 320 | 0.441 | 0.71 | |
| 0.1 | 1.9 | 12.5 | 500 | 0.441 | 0.56 | |
| 0.1 | 1.9 | 15.0 | 720 | 0.441 | 0.39 | |
| 0.1 | 1.9 | 17.5 | 980 | 0.441 | 0.22 | |
| 0.1 | 1.9 | 20.0 | 1280 | 0.441 | 0.06 | |

**Table 2**
(Continued)

| $\frac{M_{proj}}{M_{targ}}$ | $\delta$ | $\frac{v_{imp}}{v_{esc}}$ | $Q_R$ | $\Omega$ | $\frac{M_{lr}}{M_{tot}}$ | $Q_{RD}^\star$ |
|---|---|---|---|---|---|---|
| 0.1 | 2.0 | 1.0 | 3.2 | 0.408 | 1.0 | |
| 0.1 | 2.0 | 2.0 | 12.8 | 0.408 | 0.99 | |
| 0.1 | 2.0 | 4.0 | 51.2 | 0.408 | 0.95 | |
| 0.1 | 2.0 | 6.0 | 115 | 0.408 | 0.89 | |
| 0.1 | 2.0 | 8.0 | 205 | 0.408 | 0.8 | 671 |
| 0.1 | 2.0 | 10.0 | 320 | 0.408 | 0.71 | |
| 0.1 | 2.0 | 12.5 | 500 | 0.408 | 0.58 | |
| 0.1 | 2.0 | 15.0 | 720 | 0.408 | 0.42 | |
| 0.1 | 2.0 | 17.5 | 980 | 0.408 | 0.29 | |
| 0.1 | 2.0 | 20.0 | 1280 | 0.408 | 0.09 | |
| 0.1 | 2.2 | 1.0 | 3.2 | 0.354 | 1.0 | |
| 0.1 | 2.2 | 2.0 | 12.8 | 0.354 | 0.99 | |
| 0.1 | 2.2 | 4.0 | 51.2 | 0.354 | 0.95 | |
| 0.1 | 2.2 | 6.0 | 115 | 0.354 | 0.9 | |
| 0.1 | 2.2 | 8.0 | 205 | 0.354 | 0.79 | 643 |
| 0.1 | 2.2 | 10.0 | 320 | 0.354 | 0.71 | |
| 0.1 | 2.2 | 12.5 | 500 | 0.354 | 0.51 | |
| 0.1 | 2.2 | 15.0 | 720 | 0.354 | 0.36 | |
| 0.1 | 2.2 | 17.5 | 980 | 0.354 | 0.19 | |
| 0.1 | 2.2 | 20.0 | 1280 | 0.354 | 0.14 | |
| 0.1 | 2.4 | 1.0 | 3.2 | 0.311 | 1.0 | |
| 0.1 | 2.4 | 2.0 | 12.8 | 0.311 | 0.99 | |
| 0.1 | 2.4 | 4.0 | 51.2 | 0.311 | 0.96 | |
| 0.1 | 2.4 | 6.0 | 115 | 0.311 | 0.9 | |
| 0.1 | 2.4 | 8.0 | 205 | 0.311 | 0.81 | 717 |
| 0.1 | 2.4 | 10.0 | 320 | 0.311 | 0.71 | |
| 0.1 | 2.4 | 12.5 | 500 | 0.311 | 0.59 | |
| 0.1 | 2.4 | 15.0 | 720 | 0.311 | 0.42 | |
| 0.1 | 2.4 | 17.5 | 980 | 0.311 | 0.32 | |
| 0.1 | 2.4 | 20.0 | 1280 | 0.311 | 0.2 | |
| 0.1 | 3.0 | 1.0 | 3.2 | 0.222 | 1.0 | |
| 0.1 | 3.0 | 2.0 | 12.8 | 0.222 | 0.99 | |
| 0.1 | 3.0 | 4.0 | 51.2 | 0.222 | 0.96 | |
| 0.1 | 3.0 | 6.0 | 115 | 0.222 | 0.89 | |
| 0.1 | 3.0 | 8.0 | 205 | 0.222 | 0.82 | 713 |
| 0.1 | 3.0 | 10.0 | 320 | 0.222 | 0.7 | |
| 0.1 | 3.0 | 12.5 | 500 | 0.222 | 0.59 | |
| 0.1 | 3.0 | 15.0 | 720 | 0.222 | 0.41 | |
| 0.1 | 3.0 | 17.5 | 980 | 0.222 | 0.33 | |
| 0.1 | 3.0 | 20.0 | 1280 | 0.222 | 0.19 | |
| 0.1 | 4.0 | 1.0 | 3.2 | 0.144 | 1.0 | |
| 0.1 | 4.0 | 2.0 | 12.8 | 0.144 | 0.99 | |
| 0.1 | 4.0 | 4.0 | 51.2 | 0.144 | 0.96 | |
| 0.1 | 4.0 | 6.0 | 115 | 0.144 | 0.9 | |
| 0.1 | 4.0 | 8.0 | 205 | 0.144 | 0.82 | 675 |
| 0.1 | 4.0 | 10.0 | 320 | 0.144 | 0.72 | |
| 0.1 | 4.0 | 12.5 | 500 | 0.144 | 0.57 | |
| 0.1 | 4.0 | 15.0 | 720 | 0.144 | 0.43 | |
| 0.1 | 4.0 | 17.5 | 980 | 0.144 | 0.26 | |
| 0.1 | 4.0 | 20.0 | 1280 | 0.144 | 0.12 | |
| 0.1 | 5.0 | 1.0 | 3.2 | 0.103 | 1.0 | |
| 0.1 | 5.0 | 2.0 | 12.8 | 0.103 | 0.99 | |
| 0.1 | 5.0 | 4.0 | 51.2 | 0.103 | 0.96 | |
| 0.1 | 5.0 | 6.0 | 115 | 0.103 | 0.9 | |
| 0.1 | 5.0 | 8.0 | 205 | 0.103 | 0.83 | 663 |
| 0.1 | 5.0 | 10.0 | 320 | 0.103 | 0.72 | |
| 0.1 | 5.0 | 12.5 | 500 | 0.103 | 0.52 | |
| 0.1 | 5.0 | 15.0 | 720 | 0.103 | 0.38 | |
| 0.1 | 5.0 | 17.5 | 980 | 0.103 | 0.26 | |
| 0.1 | 5.0 | 20.0 | 1280 | 0.103 | 0.13 | |

**Table 2**
(Continued)

| $\frac{M_{proj}}{M_{targ}}$ | $\delta$ | $\frac{v_{imp}}{v_{esc}}$ | $Q_R$ | $\Omega$ | $\frac{M_{lr}}{M_{tot}}$ | $Q_{RD}^\star$ |
|---|---|---|---|---|---|---|
| 0.1 | 10.0 | 1.0 | 3.2 | 0.037 | 1.0 | |
| 0.1 | 10.0 | 2.0 | 12.8 | 0.037 | 0.99 | |
| 0.1 | 10.0 | 4.0 | 51.2 | 0.037 | 0.96 | |
| 0.1 | 10.0 | 6.0 | 115 | 0.037 | 0.9 | |
| 0.1 | 10.0 | 8.0 | 205 | 0.037 | 0.82 | |
| 0.1 | 10.0 | 10.0 | 320 | 0.037 | 0.7 | 653 |
| 0.1 | 10.0 | 12.5 | 500 | 0.037 | 0.58 | |
| 0.1 | 10.0 | 15.0 | 720 | 0.037 | 0.41 | |
| 0.1 | 10.0 | 17.5 | 980 | 0.037 | 0.21 | |
| 0.1 | 10.0 | 20.0 | 1280 | 0.037 | 0.11 | |

**Note.** Because the impact geometry does not matter in these cases, the final column instead lists the normalized spin rate, $\Omega$, which is related to the normalized distance $\delta$, through Kepler's third law: $\Omega^2 = \frac{4}{3\delta^3}$.

## Appendix B
## Close-in Planetary Satellites

In Table 3, we show a compiled list of all solar system satellites with measured densities and normalized orbital distance below 10.

**Table 3**
List of Planetary Satellites with Measured Densities ($\rho$) and a Normalized Orbital Distance ($\delta$) Less than 10

| Planet | Satellite | $\rho$ | $d/R$ | $\delta$ |
|---|---|---|---|---|
| Neptune | Naiad | 0.8 | 1.96 | 1.54 |
| Jupiter | Metis | 0.9 | 1.83 | 1.61 |
| Jupiter | Adrastea | 0.9 | 1.85 | 1.62 |
| Saturn | Daphnis | 0.28 | 2.34 | 1.73 |
| Neptune | Despina | 1.03 | 2.13 | 1.83 |
| Neptune | Thalassa | 1.23 | 2.03 | 1.85 |
| Uranus | Ophelia | 0.87 | 2.12 | 1.87 |
| Saturn | Pan | 0.4 | 2.29 | 1.92 |
| Saturn | Atlas | 0.41 | 2.36 | 1.99 |
| Uranus | Cressida | 0.7 | 2.44 | 2.0 |
| Saturn | Prometheus | 0.46 | 2.39 | 2.09 |
| Mars | Phobos | 1.86 | 2.77 | 2.16 |
| Uranus | Cordelia | 1.79 | 1.96 | 2.2 |
| Saturn | Pandora | 0.51 | 2.43 | 2.2 |
| Jupiter | Amalthea | 1.01 | 2.59 | 2.37 |
| Neptune | Galatea | 1.38 | 2.52 | 2.38 |
| Saturn | Epimetheus | 0.62 | 2.6 | 2.52 |
| Saturn | Janus | 0.64 | 2.6 | 2.54 |
| Saturn | Methone | 0.31 | 3.34 | 2.56 |
| Neptune | Larissa | 1.03 | 2.99 | 2.56 |
| Saturn | Pallene | 0.25 | 3.65 | 2.61 |
| Saturn | Aegaeon | 0.54 | 2.88 | 2.65 |
| Jupiter | Thebe | 0.9 | 3.17 | 2.79 |
| Saturn | Mimas | 1.15 | 3.19 | 3.79 |
| Neptune | Proteus | 1.03 | 4.78 | 4.09 |
| Saturn | Helene | 0.29 | 6.48 | 4.87 |
| Mars | Deimos | 1.46 | 6.93 | 4.99 |
| Uranus | Miranda | 1.18 | 5.12 | 5.0 |
| Saturn | Enceladus | 1.61 | 4.09 | 5.44 |
| Saturn | Tethys | 0.98 | 5.07 | 5.71 |
| Uranus | Ariel | 1.54 | 7.53 | 8.02 |
| Jupiter | Io | 3.53 | 6.03 | 8.36 |
| Saturn | Dione | 1.48 | 6.49 | 8.37 |

**Note.** We do not list density uncertainties here, although they can be relatively large for the smallest satellites on this list, owing to large uncertainties in their masses and/or sizes. $d/R$ is the satellite's mean orbital distance (semimajor axis) expressed in planet radii. These numbers are taken or derived from JPL's list of planetary satellite physical and orbital parameters, which are based on the following publications: R. A. Jacobson & V. Lainey (2014), R. A. Jacobson (2014, 2022), M. Brozović et al. (2020), and B. A. Archinal et al. (2018).

### ORCID iDs

Harrison Agrusa https://orcid.org/0000-0002-3544-298X
Patrick Michel https://orcid.org/0000-0002-0884-1993